\documentclass[preprint]{aastex}
\usepackage{amssymb}

\shorttitle{sassad}
\shortauthors{}
\begin{document}
\title{Spark Model for Pulsar Radiation Modulation Patterns}
\author{J.A. Gil, M. Sendyk}
\affil{J. Kepler Astronomical Center, Pedagogical University, Lubuska 2, 65-265, Zielona G\'ora, Poland}
\email{jag@astro.ca.wsp.zgora.pl} 

\def\RS{_{\scriptstyle RS}}
\def\be{\begin{equation}}
\def\ee{\end{equation}}
\def\ni{\noindent}
\def\st{\stackrel{<}{\sim}}
\def\gb{\gamma_{\scriptstyle_b}}
\def\gp{\gamma_{\scriptstyle_p}}
\def\bc{\begin{center}}
\def\ec{\end{center}}
\def\bi{\begin{itemize}}
\def\ei{\end{itemize}}
\def\sti{\scriptstyle_{T_i}}
\def\gi{\gamma_{\scriptstyle_i}}
\def\pms{pulsar magnetosphere}

\begin{abstract}
   A non-stationary polar gap model first proposed by Ruderman \&
Sutherland (1975) is modified and applied to spark--associated 
pulsar emission at radio wavelengths. It is argued that under physical
and geometrical conditions prevailing above pulsar polar cap, highly
non-stationary spark discharges do not occur at random positions. Instead, 
sparks should tend to operate in well determined preferred regions.
At any instant the polar cap is populated as densely as possible with a
number of two-dimensional sparks with a characteristic dimension as well
as a typical distance between adjacent sparks being about the polar gap
height. Our model differs, however, markedly from its original ``hollow
cone'' version. The key feature is the quasi-central spark driven by
$\gamma-B$ pair production process and
anchored to the local pole of a sunspot-like surface magnetic field.
This fixed spark prevents the motion of other sparks towards the pole,
restricting it to slow circumferential ${\bf E}\times{\bf B}$ drift across the planes of
field lines converging at the local pole.
We argue that the polar spark constitutes the core pulsar
emission, and that the annular rings of drifting sparks contribute to conal
components of the pulsar beam. We found that the number of nested cones in the 
beam of typical pulsar should not excced three;
a number also found by Mitra \& Deshpande (1999) using a completely
different analysis.
We confront predictions of our model with
a variety of pulsar data, including mean profiles morphology and their
predicted correlations with properties of the
$P-\dot{P}$ pulsar diagram as well as detailed studies of drifting subpulses 
(Deshpande \& Rankin 1999). We demonstrate
that, if the observing geometry is known, the average profile as well as the apparent drift 
pattern are fully determined by the values of $P$ and $\dot{P}$. In the accompanying Paper II we develop
a self-consistent theory of coherent pulsar radio emission based on the modified polar gap
model explored in this paper.
\end{abstract}


\section{Introduction}

Pulsar radiation is believed to originate within a bundle of open
magnetic field lines, along which a significant part of a rotation
induced potential drop can occur. This potential drop accelerates charged
particles supplied by the polar cap area, which is a region of the neutron
star surface directly connected to the interstellar medium via magnetic
field lines. Two types of polar cap activity models have been proposed
so far. In the first type, called the free-flow or stationary models
\citep{saf78,as79,a81}, the
charged particles flow freely from the polar cap surface and accelerate
within a scale height of about one stellar radius $R\approx 10^6$~cm, due
to
the potential drop resulting from the curvature of field lines and/or
inertia of outstreaming particles. In the second type, called gap models
or non-stationary models (Sturrock 1971; Ruderman \& Sutherland 1975 - RS75 henceforth; Cheng
\& Ruderman 1977, 1980), the free outflow from the polar cap surface is strongly
impeded, which leads to the formation of an empty gap just above the polar cap. The
high potential drop across the gap is discharged by the photon-induced pair
creation in the strong and curved magnetic field. This breakdown of the
polar gap stabilizes its height to about one mean free path for the
$\gamma-B$ pair production process, which has to be smaller than the polar
cap radius $r_p\approx 10^4P^{-1/2}$~cm, where $P$ is pulsar period in
seconds. The charges of opposite signs are
accelerated in opposite directions to energies exceeding $2m_ec^2$ and a
cascade avalanche develops. The polar gap breaks in the form of a number of
isolated, short-lived discharge tubes called sparks
\citep{cr77,b82,fr82}, which can be naturally interpreted in terms of subpulse-associated
plasma columns, modulating the pulsar radiation spatially on the subpulse
time-scales ($\sim 1$~ms). 

The plan of the rest of this is as paper follow. In section 2 we revise the RS75 model in
such a way that it is no longer the hollow cone type model. We assume that
the surface magnetic field is non-dipolar, with a typical radius of curvature
much smaller than the neutron star radius ${\cal R}<10^6$ \citep{bah83,rom90,rud91a,k91,a93,cr93,mkb99}. 
We also assume
that the planes of the magnetic fields lines should tend to converge at
the local pole, introducing some degree of axial symmetry like in the case of ``sun-spot'' 
field configurations \citep[][see also Appendix]{cr93,gm99}. We follow the RS75
idea that the high potential drop above the polar cap is discharged via a
number of localized spark filaments and argue that one of them should be
anchored to the local surface pole, while others should perform more or less
ordered circumferential motion around it due to the well-known ${\bf E}\times{\bf B}$
drift. However, we take into account reduction of the vacuum 
accelerating potential drop (eq. [\ref{deltamax}]) due to spark development (eq. [\ref{calef}]) , which reduces 
also the apparent drift rate (eqs. [\ref{vperp}], [\ref{pehat}] and [\ref{N}]). 
We argue that both the characteristic dimension as
well as the typical distance between sparks should be about the RS75 gap
height (eq. [\ref{h}]). We use this result in section 3, where we propose that polar spark is
associated with the core pulsar emission, while concentric rings formed by drifting
sparks correspond to nested conal components of the ``non-hollow cone'' radiation
pattern. We calculate the number of cones in the pulsar beam and, since it depends
mainly on the values of basic pulsar parameters $P$ and $\dot{P}$, we
examine correlations of number of profile components and profile types with
properties of
standard $P-\dot{P}$ diagram. The degree of correlations is high, supporting
our picture. In section 4 we revise the RS75 subpulse drift model and 
apply it to a number of pulsars with drifting subpulses. 
In section 5 we incorporate the behaviours of drifting subpulses into the
core/cone morphological model of pulsar proposed and developed by \citet{r83}.
We discuss both general and specific implications of our model in section 6.
In the Appendix we briefly discuss evidence of nondipolar surface magnetic field from 
PSR J2144-3933, the recently
reported 8.5-s pulsar.

Despite a number of well-known problems with the neutron star crust binding energy 
\citep[see][for short review]{xqz99}, the recent papers concerning pulsars with drifting 
subpulses PSR 0943+10 \citep{dr99} and PSR 0031$-$07 \citep{vj99} strongly suggest that the RS75-type 
vacuum gap does exist above the polar cap. Motivated by these evidences, \citet{xqz99}
have recently  argued that the binding 
energy problem can be completely solved if pulsars (at least those with
drifting subpulses) represent bare strange stars, which in 
many aspects are indistinguishable from neutron  stars. 

Sparks proposed in the RS75 model as primary sources of the subpulse-associated
plasma columns have been criticized, mainly due to short dynamical time
scales ($\sim 10~\mu$s) as compared with subpulse time scales ($\sim
1$~ms). To account for the subpulse emission, they would have to repeat at
approximately the same place for a time long relative to their lifetime. No clear explanation for such kind of a surface memory was
proposed by RS75. In this paper we propose a modification of
the RS75 polar gap model and confront the predictions of the revised model
with the much larger body of present day observational data, concerning
both subpulses in single pulses and average profiles.
 
New ingredients of our modified RS75 model include: (a) estimation of a characteristic
spark dimension ${\cal D}\approx h$ (eqs. [\ref{h}] and [\ref{calde}]), (b) introduction of quasi-central spark anchored to the local 
pole of surface magnetic pole (thus our model is no longer of 
a hollow-cone type); (c) reduction of gap potential drop corresponding to subpulse emission by a filling
factor ${\cal F}<1$ (eq. [\ref{calef}]). The critical assumption we need to adopt is an existence of a local
magnetic pole (different from the global dipole) at the polar cap. This is necessary
for two reasons: (i) a radius of curvature of surface magnetic field is small enough
to drive and exponentiate spark discharges in the first place; (ii) the spark fixed at the local pole
warrants a persistence of spark arrangement in the form of a quasi-annular pattern leading to 
nested cone structure of the pulsar beam. Such a multiconal organisation of pulsar beams was first suggested by
Rankin (1983, 1993a, b), supported by \citet{gks93} and confirmed ultimately 
by \citet{md99}. The  recently reported pulsar with longest 
period $P=8.5$~s \citep{ymj99} suggests that the surface magnetic field in pulsars has a 
sunspot-like configuration \citep[][see also Appendix]{cr93,gm99}. We argue that one of the local poles is located 
at the \citet{gj69}  polar cap, and assume that pulsar radio emission originates on those
multipolar surface field lines which reconnect with dipolar ones in the radio emission region.

\section{Polar gap sparking discharge}

In this section we discuss the spark reappearance and spark characteristic
dimension problems.
As described by RS75, the polar gap discharges through a number of
localized sparks, separated from each other by a distance approximately
equal to the gap height 
\be
h=5\cdot 10^3B_{12}^{-4/7}{\cal R}_6^{2/7}P^{3/7}~{\rm cm} ,\label{h}
\ee
where $P$ is pulsar period in seconds, $B_{12}=B_s/10^{12}$~G is the surface magnetic field in units
$10^{12}$ Gauss and ${\cal R}_6={\cal R}/10^6$~cm in the radius of
curvature of surface field in units of $R=10^6$~cm. We assume that surface magnetic field $B_s$ is
highly multipolar, that is 
$B_s=b\cdot B_d$, where $B_d=3.2\cdot 10^{19}(P\cdot\dot{P})^{1/2}$~G and a dimensionless factor $b\gg 1$. It seems
natural to assume the range of dimensionless curvature radii from within the range 
$0.1\stackrel{<}{\sim}{\cal R}_6<1.0$ \citep[][see Appendix]{gm99}.
Each spark 
develops exponentially
until the plasma density $\rho$ reaches a value  close to the corotational
Goldreich-Julian (1969) density $\rho_{GJ}$, 
 screening the gap locally. The potentional drop within the spark filament
is roughly $\Delta V=(1-\rho/\rho_{GJ})\Delta V_{max}$, where
$\rho<\rho_{GJ}$ and the maximum RS75 potential drop
\be
\Delta V_{max}=1.7\cdot 10^{12}{\cal R}_6^{4/7}B_{12}^{-1/7}P^{-1/7}~{\rm V} .
\label{deltamax}
\ee
The spark exponentiation stops abruptly when $\rho$ approaches $\rho_{GJ}$. The potential drop $\Delta V$ is reduced
to a value slightly below the treshold for $\gamma-B$ pair production
\be
\Delta V_{min}=\gamma_{min}m_ec^2/e~, \label{Vmin}
\ee
where
$$
\gamma_{min}=(2m_ec^2{\cal R}/c\hbar)^{1/3}\approx 3.3{\cal
R}_6^{1/3}10^5 .
$$
In this expression $\hbar$ is the Planck constant, $e$ and $m_e$ is the electron charge
and mass, respectively, and $c$ is the speed of light .
Note that unlike equation~(\ref{deltamax}), equations~({\ref{Vmin}) and (\ref{calef}) 
represent a necessary but not sufficient
condition for magnetic pair production within the polar gap.
One can therefore introduce a filling factor
\be
{\cal F}=\frac{\Delta V_{min}}{\Delta V_{max}}\approx 0.1{\cal
R}_6^{-5/21}B_{12}^{1/7}P^{1/7}\ll 1~,\label{calef}
\ee
which determines a saturation stage at which the spark plasma begins to leave the
gap. We will call it the ``gap emptying stage'', in which spark plasma is dense enough 
to contribute efficiently to the mechanism of generation of coherent radio emission
(see Paper II). When the spark plasma
leaves the gap with a speed $v\stackrel{<}{\sim}c$, the potential drop beneath the spark
 rapidly grows and
it should exceed the threshold value for $\gamma-B$ pair creation before the
bottom of the spark filament reaches the top of the gap. It is obvious
that the gap
emptying stage does not begin simultaneously in adjacent sparks, each
screening the area within a distance $h\sim 5\cdot 10^3P^{3/7}$~cm. Therefore, the returning
potential drop will be larger beneath the leaving spark plasma than around
it and the very next cascade should be initiated and developed in
approximately the same place as the previous one, provided that the spark
plasma does not move fast in any direction during the exponentiation time.
Such motions are prevented by a putative spark anchored to the local pole. This
will be discussed in more detail later in the paper. 

In order to avoid
confusion, we will now give a short summary of the remaining part of this
section, so the reader knows what to expect. First we estimate a
characteristic spark dimension ${\cal D}_\perp$ as seen by observer's
line-of-sight (perpendicular to the planes of field lines converging at the
local surface pole). Next, we estimate a characteristic dimension ${\cal
D}_\|$ resulting from a rapid spread of spark avalanche within the
planes of field lines (parallel). Then we notice that since ${\cal
D}_\|\approx{\cal D}_\perp\sim h$, a mechanism must exist broadening an avalanche also in
the perpendicular direction. We invoke a photon splash known from
literature. All in all we argue that sparks should be
a two dimensional entities on the polar cap, with a characteristic dimension
${\cal D}\sim{\cal D}_\perp\approx{\cal D}_\|$, as well as a separation of adjacent 
sparks, being about the polar gap
eight $h\sim 5\cdot 10^3P^{3/7}$~cm in typical pulsars.

Interpreting the width of subpulses as radio emission from plasma
columns flowing along dipolar field lines connected to sparks, one can
estimate the fraction $({\cal D}/r_p^d)^2$ of the polar cap area filled by a spark
being in the range $10^{-1}$ to $10^{-2}$ \citep{cr77}, where ${\cal D}$ is a characteristic
spark dimension referred to a dipolar polar cap. We can form a dimensionless parameter
\be
r_p^d/h_{RS}=5\cdot 10^4{\cal R}_6^{-2/7}\dot{P}^{2/7}P^{-9/14} ~,\label{erdepe}
\ee
where $r_p^d=10^4P^{-1/2}$~cm is the Goldreich-Julian (1969) 
polar cap radius and $h_{RS}=h$~ (eq. [\ref{h}]) is the RS75
polar gap height with a dipolar component of surface magnetic field 
$B_{12}=B_{12}^d$ (in units of $10^{12}$ Gauss). For a typical pulsar with $P\approx 1s,
\dot{P}=10^{-15}$ and ${\cal R}_6^{2/7}\approx 0.7$ this equation gives
$h_{RS}\approx 0.2~r_p^d=2\cdot 10^3$~cm. On the other hand
$<{\cal D}/r_p^d>=\sqrt{0.05}\approx 0.2$ or ${\cal D}\sim 0.2~r_p^d$. By comparison one
can conclude that ${\cal D}\approx h_{RS}$. This means  that the characteristic dimension of
the subpulse-associated plasma column projected onto the polar cap along
dipolar field lines is approximately equal to the RS75 polar gap height $h$
(eq.~[\ref{h}]). One should emphasize, however, that this is only observational
constraint corresponding to the dimension along the line-of-sight, as there is 
no direct information available about the other dimension. However, \citet{dr99} using their
carthographic transform technique applied to good quality single pulse data of PSR 0943+10, clearly
demonstrated that sparks are two-dimensional, approximately circular entities related to drifting subpulses,
as originally proposed by RS75 (see also Fig.~1 in this paper for illustration).

In the case of the dipolar surface magnetic field ${\cal D}$ can be directly
interpreted as a spark characteristic dimension (at least in the direction
along the line-of-sight). In the actual pulsar however, both the polar cap
radius and the polar gap height have to be modified to
include higher multipoles dominating the global dipole magnetic field at the
polar cap surface. As a result, equation~(\ref{erdepe}) should be replaced by
\be
r_p/h=b^{1/14}r_p^d/h_{RS} ,\label{erpe}
\ee
where $r_p=b^{-1/2}\cdot r_p^d$ and
$B_{12}=b\cdot B_{12}^d$. As one can see, the actual ratio $r_p/h$
does not differ by more than a factor of two from the canonical value
(eq.~[\ref{erdepe}]), even if the actual surface magnetic field is $10^4$ times stronger
than the dipolar component inferred from the pulsar slow down rate. This
simply means that the angular ratio of the spark-associated plasma column to
the open field lines region is preserved down to the polar cap, no matter
how complicated the surface magnetic field is. Treating $b$ as a constant in equation (\ref{erpe}) means that
we assume that the actual surface magnetic field evolves similarly to that of purely 
dipolar field. As demonstrated recently by \citet{mkb99}, such assumption is quite well justified.

One can attempt to estimate the spark dimension using an independent
argument. The number density of the $e^-e^+$ pairs in the sparking
avalanche should develop exponentially with time. To estimate precisely the
spark exponentiation time $\tau$, that is the characteristic time-scale
after which the spark charge density reaches the
corotational value $n_{GJ}$ which screens the gap (where $n_{GJ}$ is the
Goldreich-Julian (1969) number density), one would require a detailed
physical model for pair formation and spark development within the gap region. So far such
a model does not exist and we have to use arguments based on a general picture
of pair creation in strong curved magnetic fields (Erber 1964, RS75)
and dimensional analysis. Let us notice that the exponentiation time-scale
$\tau$ should be approximately proportional to the radius of curvature ${\cal R}$ of the
surface magnetic field. In fact, copious magnetic pair production within the gap
requires a large perpendicular component of the magnetic field $B_\perp\sim
hB/{\cal R}$. For a given $h$ and $B$, the curvature photon has to travel a
distance $l\leqslant h$ to reach a value of $B_\perp$ high enough to produce a
pair. The smaller the radius of curvature ${\cal R}$, the smaller the
distance $l$ and, in consequence, the shorter the sime-scale $\tau$. 
Thus, a natural spark exponentiation time-scale is
\be
\tau\approx{\cal R}/c .\label{tau}
\ee
For a small radius of curvature ${\cal R}\approx 3\cdot 10^5$~cm (see Appendix) this is about
10 $\mu$s, which is equal to the value obtained by independent arguments in
the RS75 model (see also Beskin 1982 and Asseo \& Melikidze 1998).

If the surface magnetic field is not extremely tangled and posseses some
degree of quasi-axial symmetry, like in the case of sunspot-like configuration,
 the sparks will develop (at least
initially) in the form of thin plane sheets, following the planes of 
field lines converging at the local pole. In fact, the spark plasma is subject to
fast parallel drift motion towards a local pole (in the direction opposite to field lines
curvature) with a speed
\be
v_{\|}\approx ch/{\cal R} \label{vpar}
\ee
(Cheng \& Ruderman 1977, 1980; Filippenko \& Radhakrishnan 1982). Thus,
during the exponentiation time $\tau\sim{\cal R}/{c}$ the spark will cover a
distance ${\cal D}_\|=v_\|\cdot\tau\approx(hc/{\cal R})({\cal R}/c)=h$.
This can also be considered as a characteristic spark dimension (in the planes
of field lines converging towards a local magnetic pole). As one can see,
${\cal D}_\|$ is equal to the perpendicular spark dimension ${\cal D}_\perp$
inferred from the subpulse widths analysis. Therefore, a mechanism has to
exist which spreads an initially thin and plane discharge into a
two-dimensional entity with a characteristic dimension
\be
{\cal D}\approx h ,\label{calde}
\ee
where $h$ is the gap height (eq. [\ref{h}]). One such mechanism, called a ``photon
splash'' effect has been proposed by Cheng \& Ruderman (1977). It occurs
when a very energetic electron (positron) impacts the pulsar surface. As a
result, at least one secondary high-energy ($>1$~MeV) gamma ray is emitted.
Since this emission is undirectional, the photon splash can effectively
blow a
spark virtually in all directions within the limit of a free path for
$\gamma-B$ pair production, which is approximately equal to the RS75 gap
height $h$. Thus, a fully developed spark should be approximately circular in shape with 
diameter ${\cal D}\approx h$ (eq. [\ref{calde}]), as demonstrated recently by \citet{dr99} analyzing
the case of drifting subpulses in PSR 0943+10 (see also Fig.~1 in this paper).

In principle, the spark plasma is subject to two drift motions:
fast ``parallel'' drift along the planes of field lines with a velocity
described by equation~(\ref{vpar}) and a much slower ``perpendicular'' ${\bf E}\times{\bf B}$ drift
(RS75) across the planes of field lines with a velocity
\be
v_\perp=c\Delta E/B_s={\cal F}\frac{2\pi}{P}\frac{h^2}{r_p(1-s)} ,\label{vperp}
\ee
where $r_p\approx 10^4P^{-1/2}$~cm is the polar cap radius, 
$\Delta E={\cal F}(2\pi/cP)\left(h^2/r_p(1-s)\right)\cdot B_s$ is 
the gap electric field perpendicular to the surface magnetic field $B_s$, 
${\cal F}$ is the filling factor corresponding to
the spark termination stage (eq.~[\ref{calef}]) and $s=d/r_p$ is the mapping 
parameter $(0<s<1)$ of field lines associated with a spark operating at a distance $d$ from
the pole. Typically $v_\|\gg v_\perp$, so ${\bf E}\times{\bf B}$ drift is not expected
to influence the spark shape. It will however, cause the slow motion of fully developed
sparks around the local centre of axial symmetry, determined  by the pole of sunspot-like surface
magnetic field.

Let us consider an avalanche discharge which by chance occured close to the
local pole of the multipolar surface magnetic field (which is quite likely to happen
given a large number of energetic $\gamma$-quanta penetrating
the gap region; see RS75). Due to both parallel and perpendicular drift motions as
well as a putative quasi-axial symmetry at the pole, this discharge will
very soon
anchor itself to the polar area and form a fixed spark 
(${\bf E}\times{\bf B}$ circulating around ``itself''). 
This should happend independently of any details of polar gap discharge.
In fact, at the very begining sparks will rush towards the local pole due to quasi-axial symmetry of field line
planes. Once a spark reaches a pole it begins to circulate around itself (due to ${\bf E}\times{\bf B}$ drift). We
will interpret this fixed polar spark as the source of plasma column related to the
core pulsar emission (Rankin 1983, 1990). Other sparks, which can form at a screening 
distance $h\sim h_{RS}$ from the polar spark and from each other (see Fig. 1 for illustration), will perform circumferential
${\bf E}\times{\bf B}$  drift around the pole. It is natural to interpret plasma
columns associated with these sparks as sources of the conal pulsar emission (Rankin 1983, 1990, 1993a, b).

To summarize this section; when a sparking discharge begins at some point on
the polar cap, the gap potential drop rapidly falls below the
$\gamma-B$ pair formation threshold, which should inhibit another discharge
within a distance of about $h$ (eq.~[\ref{h}]). The fixed polar spark thus restricts degrees of
freedom of other sparks to circumferential motion around the pole, as they
 cannot approach the polar one within a screening
distance $h$.
The existence of the polar spark is therefore crucial for our model.
Within the $\gamma-B$ pair production mechanism of
gap discharge, it requires that the polar field lines are sufficiently
curved (like in a ``sunspot'' configuration). However, one cannot exclude
that the actual gap magnetic field is fully axially-symmetric (like the
case of a star-centered dipole or quadrupole). In such a case the polar field lines
are not curved enough to maintain a cascade of $\gamma-B$ pair production.
Zhang \& Qiao (1998) have recently proposed an alternative two-photon
pair production process, which may perhaps account for the formation of
a polar spark in a fully symmetric field configuration, in which there is not
enough curvature to drive $\gamma-\beta$ pair production at the pole. 


\vskip 12pt~
\scalebox{0.85}
{\includegraphics{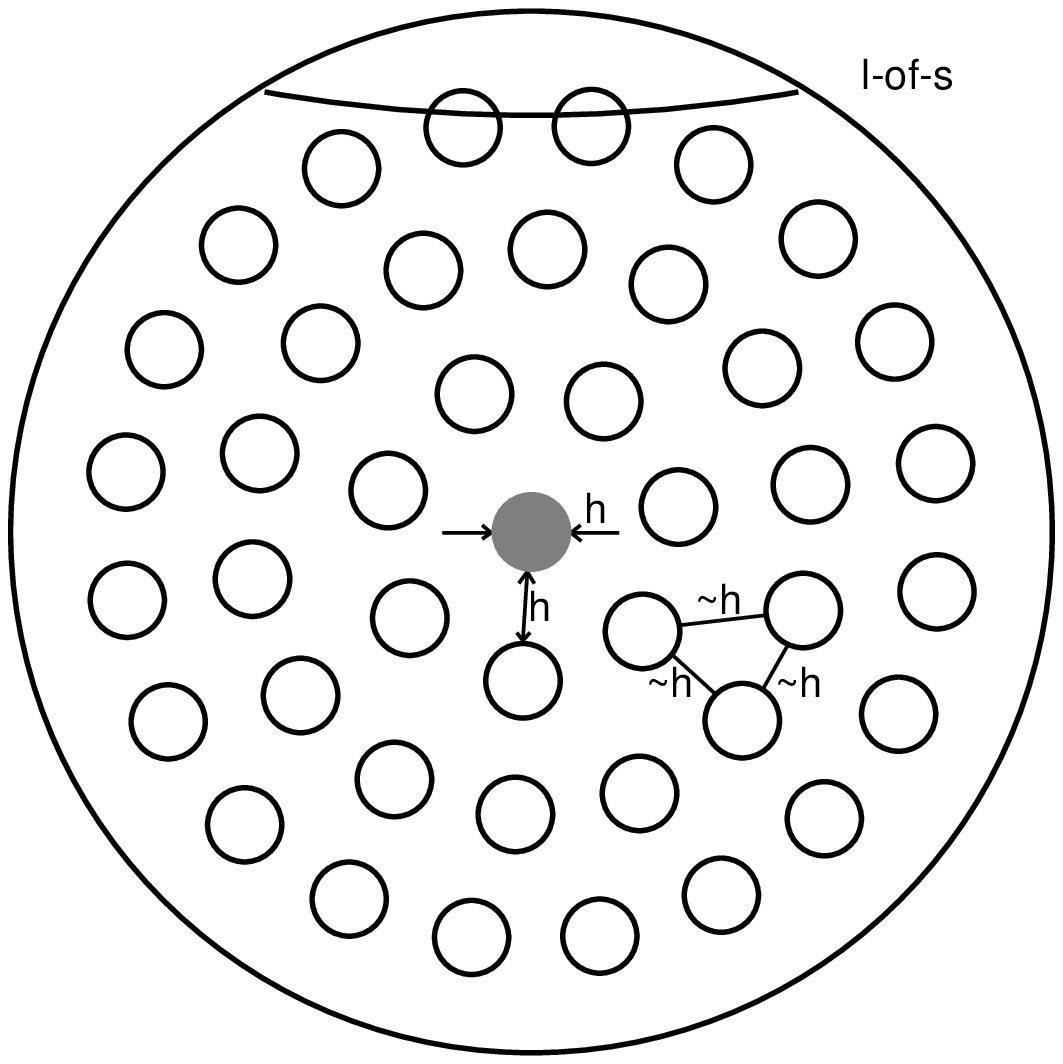}}~\figcaption[fig1.eps]{The illustration of instantaneous arrangement of 
a number of equi-distant sparks on
the polar cap of PSR B0943+10. The shadowed spark is 
anchored to the local pole of the surface magnetic pole while other sparks
perform a slow ${\bf E}\times{\bf B}$ drift around it. The line-of-sight
cuts through the outermost ring of 20 equaly spaced sparks at the impact
angle $\beta=-8^\circ$ from the pole (for comparison see the map of 20
subpulse beams in Deshpande and Rankin, 1999).
Both the HPBW of each spark and a distance between adjacent sparks is about $h=r_p/a\sim 17$~m,
where the polar cap radius $r_p$ is about 110 m and the complexity parameter $a\sim 6.5$
(see section 4.1 for explanations).
\label{fig1}}~\vskip 2pt

\section{Polar cap structure and mean pulsar beams}

As argued in previous section, the actual pulsar polar cap should be
populated by a number of sparks with a characteristic dimension ${\cal
D}\approx h$, separated from each other also by about $h$, where $h$ is the
actual gap height close to the value given by RS75 (eq.~[\ref{h}]). One spark should be always active at the
local surface magnetic pole, and other sparks should perform more or less
ordered,
circular ${\bf E}\times{\bf B}$ drift motion around the polar spark.
In this sense our model differs markedly from the RS75 hollow-cone version in
which sparks occupied only an outermost parts of the polar cap.
Since the polar spark prevents the motion of other sparks towards the pole, the
drifting sparks should form on the average a multi-ring structure centered on
the polar spark. These rings are not expected to be perfectly circular, as we
assume only quasi-axial symmetry of field line planes converging at the pole
 of a surface magnetic field.
 It is natural to assume that plasma supplied by sparks to
the
magnetosphere above the polar gap will eventually give rise to the subpulse
emission (see Paper II) at altitudes of a few percent of the light-cylinder radius
$R_{LC}=cP/2\pi$ (Cordes 1992, Kijak \& Gil 1997, 1998). The subpulses
associated with the polar spark should be longitude-stationary 
and
constitute the so called ``core component'' in the mean pulsar profile,
while other subpulses may demonstrate an organized drift from one pulse to
another and contribute to the so called ``conal components'' in the mean
pulsar profile. In section 5 we discuss distinction
between core and conal profile components in more detail. 

We would like to emphasize thet the ring structure of the polar cap mean energy distribution
is a consequence of (i) fixed spark at the local pole, and (ii) screening distance
between adjacent sparks approximately equal to spark dimension (Fig.~1). This structure
will result in a specific organization
of the mean pulsar emission pattern, consisting of a narrow core beam
surrounded by
a number of nested conal beams. Simulations of pulsar emission within the framework
of the above model seem to reproduce the observational properties of both
single pulses and mean profiles quite well (Gil et al. 1995, Gil \& Krawczyk 1996, 1997). 
Such a nested cone structure of pulsar beams
have bean suggested by many authors (e.g Rankin 1993; Gil, Kijak \&
Seiradakis 1993; Kramer et al. 1994). Alternatively, the patchy beam model has also been proposed
to explain variety of pulsar profiles (Lyne \& Manchester 1988). 
However, as demonstrated by Gil \& Krawczyk (1996), the patchy beam model is inconsistent
with the observed frequency evolution of subpulses as contrasted with profile components. 
Also Mitra \& Deshpande
(1999) have recently presented strong
arguments against patchy beams or even thick hollow cone beams. Analyzing multifrequency 
pulse-width data, Mitra \& Deshpande clearly revealed a nested cone structure of pulsar beams and
found that each cone is illuminated in the form of annular ring of width being typically
about 20 percent of the overall beam radius (opening angle). This is in perfect agreement with
our ratio of spark diameter to polar cap radius ${\cal D}/r_p~\sim 0.2$ (see discussion below 
equation (\ref{erdepe})). 

Within our geometrical model of the polar cap, the complexity of pulsar emission pattern 
should just be 
determined by this ratio of 
the polar cap radius $r_p\sim 10^4P^{-1/2}$~cm to the spark characteristic dimension 
${\cal D}\sim h=5\cdot 10^3\dot{P}^{-2/7}P^{1/7}$~cm.
Following equations (\ref{erdepe}) and (\ref{erpe}) we will introduce a {\it complexity parameter}
\be
a=\frac{r_p}{h}\approx 5\cdot\dot{P}^{2/7}_{-15}P^{-9/14} ,
\label{a}
\ee
where $\dot{P}_{-15}=\dot{P}/10^{-15}$. To make it possibly independent of unknown characteristics of the actual
surface magnetic field, we have assumed that $b^{1/14}{\cal R}_6^{-2/7}\approx 2$ (which is  roughly the case
for all realistic combinations of $b>1$ and ${\cal R}_6<1$; see Appendix). 

Each spark can be modelled with a Gaussian of
half-power beam width about $h$. According to RS75, distance between HPBW points is also about $h$.
This means that the complexity parameter $a$ (eq. [\ref{a}]) determines roughly a number of equi-spaced sparks
$N_{sp}\sim a^2$ (see Fig.~1 for illustration) that can operate on the entire polar cap at any given time (see also Beskin 1992).
It is natural to assume that these sparks, rotating around fixed spark anchored to the local pole, form on the
average a system of nested cones in the pulsar beam. We additionaly assume that the fixed spark contributes to the
core pulsar emission. The number of nested cones surrounding the core beam can be roughly described by
\be
n=Int\left(\frac{a-1}{2}\right). \label{n}
\ee
(half of the Gaussian sparks that can be fit along a polar cap diameter, excluding the one fixed at the pole).
As a consequence of equations (\ref{a}) and (\ref{n}), the number of components resolved in the average pulsar profile
is approximately given by
\be
N\leq N_{max}=2\cdot n+1\approx Int(a) ,
\label{N}
\ee
depending on the observer's impact angle, where the equality holds when the
light-of-sight passes close to the axis associated with the core beam. However,
it should be stressed that profile components can be resolved only for relatively small values of
the complexity parameter $a\stackrel{<}{\sim}10$ (for $a>10$ the above equation reads $N=1$;
see section 5.3b). 

The subpulse emission in single pulses can be resolved only
if the value of $a$ is not too large. This means that subpulses 
(possibly drifting) can be detected in
pulsars with relatively small values of $a$, which occurs in older pulsars.
The radio emission of younger pulsars with shorter periods should be
predominantly amorphous with no hints of modulation on the subpulse time
scales. 

\section{Spark-associated subpulse drift}

The  ${\bf E}\times{\bf B}$ drift across the planes of
field lines converging at the local pole, responsible for creating
a nested cone structure of average pulsar beams, should manifest itself as a prominent
subpulse drift and/or periodic intensity modulation in subpulses associated
with conal components. 
The thick circles in Fig.~5 indicate presence of drifting subpulses 
in single pulses of corresponding pulsars.
As demonstrated by Rankin (1986) drifting subpulses are a purely conal
phenomenon.
It is clear from
our model why drifting subpulses tend to occur mostly in conal components
of complex profile pulsars ($M,Q,cT,D,S_d$) and never in core-single
($S_t$)
profile pulsars. First, subpulses in core components of complex profiles
cannot drift, since the corresponding spark is anchored to the local
magnetic pole. As for the core-single pulsars $(S_t)$, the lack of an apparent
subpulse drift is a consequence of unresolved subpulse beams in their pattern emission 
(see section 5.3 for explanation). However, subpulse drift is in principle possible in the conal
outriders developed at higher frequencies and in conal components of
Triple ($T$) profile pulsars \citep{r83}. 

The ${\bf E}\times {\bf B}$ drift results in a slow
circumferential motion
of sparks around the local pole of surface magnetic field. 
The spark associated plasma column completes
one rotation in a time interval
\be
\hat{P}_3=2\pi d/v_\perp=(P/{\cal F})\cdot a^2\cdot s(1-s) ,\label{pehat}
\ee
where $d=s\cdot r_p$ is distance of spark centre from the pole (the mapping parameter $s=0$ at the pole
and $s=1$ at the polar cap edge $d=r_p$), $v_\perp$ is the speed of circulation described by
equation (\ref{vperp}) and $a=r_p/h$ is the complexity parameter (eq.~[\ref{a}]). 
The number of circulating sparks which contribute to the observed drift pattern is
\be
{\cal N}=\hat{P}_3/P_3 ,
\label{N1}
\ee
where $P_3$ is the number of periods $P$ between two primary drift-bands. The rate of circulation is
${\cal D}=360^\circ P/\hat{P}_3$, which for typical pulsars is about 10 degrees per period. 
Let us note that if $n>1$ (eq. [\ref{n}]) then ${\cal N}<{\cal N}_{sp}\sim a^2$.

The above equations represent modification of RS75 drift description in two aspects, namely:
(i) they can be applied to spark at any distance $d=s\cdot r_p$
from the pole ($s=0.5$ in RS75), provided that spark diameter ${\cal D}\sim h\ll r_p-d$, and (ii) the
gap electric field is reduced by a factor ${\cal F}\sim 0.1$ (eq. [\ref{calef}]) related to 
fully developed spark plasma (${\cal F}=1$ in RS75). 
Below we explore the modified RS75 spark model in order to explain and reproduce
patterns of drifting subpulses in a number of pulsars, for which good
quality single pulse data were available.

Let us begin with a short summary. We will simulate the single pulse emission patterns of 
four pulsars by performing a number of subsequent steps: (i) determination of a
number of nested cones and a number of sparks associated with the outermost
cone (eqs. [\ref{a}] and [\ref{n}]); (ii) estimation of the mapping parameter $s$
corresponding to the locus of the outermost cone; (iii) estimation of the
approximate values of the inclination $\alpha$ and impact $\beta$ angles
using spectral and polarization information; (iv) determination of the
fundamental periodicity $\hat{P}_3$ using equation (\ref{pehat}); (v) estimation of
emission altitude; (vi) calculations of the single
pulse sequence corresponding to a system of drifting sparks
(e.g. Fig.~1); (vii) matching the calculated ond observed patterns by
fine tuning the values of $\alpha$ and $\beta$, and if possible, testing the
self-consistency of the model by calculating the expected polarization position angle
curve and comparing it with that observed. For details of the simulation technique, see \citet{gkm95}. 

\subsection{PSR B0943+10}

Recently, Deshpande \& Rankin (1999) have analysed in unprecedented
detail an extraordinarily stable drifting-subpulse pattern of PSR B0943+10 (see Fig.~2b). This
is an interesting pulsar which exhibits the so-called
even-odd modulation caused by the fast subpulse drift corresponding to the apparent periodicity 
$P_3/P\sim 2$ \citep{so75}. This even-odd modulation manifests itself by apparent secondary driftbands
of subpulses corresponding to every other pulse period. Of course, the primary driftbands
corresponding to consequtive pulse periods are not visible in this pulsar. 
Deshpande \& Rankin (1999) determined the observing geometry
corresponding to a peripheral sightline grazing the outer beam, in which
they were able to identify 20 rotating subbeams producing the apparent drift pattern.
Each spark completes one full rotation in 37 periods $P$.
Deshpande \& Rankin (1999) have found this picture generally consistent with the system
of sparks on the polar cap (RS75), with dimensions determined by the gap
height. We provide a more specific description of the corresponding spark 
system, using equations (\ref{pehat}) and
(\ref{N1}). 

The complexity parameter of this pulsar $(P=1.098\ 
{\rm s}, \dot{P}_{-15}=3.52)$ $a\approx 6.8$ (eq.~[\ref{a}]). This
implies that the mean radio emission should be arranged in 
$n=3$ nested cones around
the core beam (eq. [\ref{n}]). The geometrical model of the polar cap for this pulsar shows 7, 14 and 20
spark-associated subpulse beams corresponding to 1st, 2nd and 3rd cone,
respectively (Fig.~1). Given the grazing sightline geometry, an observer
scans the outermost cone constisting of ${\cal N}=20$ sub-beams circulating around
the magnetic axis with a linear velocity determined by equation~(\ref{vperp}) and completing
one full rotation in the time interval described by equation~(\ref{pehat}). The locus of the
outermost cone on the polar cap corresponds to the mapping parameter $s_{out}=0.875$. 
To obtain exactly the time interval $\hat{P}_3=37\cdot P$ provided by the
sofisticated analysis of Deshpande \& Rankin (1999), we just need to adopt for a filling
factor ${\cal F}=0.125$, which is perfectly consistent with ${\cal F}\sim 0.1$ expected from 
equation (\ref{calef}). This value of ${\cal F}$,
together with $a^2=46$ and $s=0.875$ gives $\hat{P}_3/P=36.8$ (eq.~[\ref{pehat}]).
Let us note that according to equation (\ref{N1}] the apparent periodicity $P_3=1.85~P$, close to 
$P_3=1.89~P$ provided by \citet{so75}.

It is worth to mention here
that the ``vacuum'' value of $\hat{P}_3$ given by RS75 is  much shorter
(they use ${\cal F}=1$ and $s=0.5$, for which values equation (\ref{pehat}) gives $\hat{P}_3\sim 11P$ if
${\cal N}=20$ and even less for ${\cal N}<20$). 
In fact, $\hat{P}%
_3/P\approx 5.6\cdot B_{12}/P^2$ in their model, which for $P\approx 1$~s
and $B_{12}\approx 2$ gives $\hat{P}_3\approx 9P$. Thus, the analysis of
drifting subpulses in PSR B0943+10 performed by Deshpande \& Rankin (1999)
strongly supports our modified RS75 model. 

To simulate single pulses of PSR B0943+10 at frequency 0.43 GHz (to be
compared with the Arecibo data presented in Fig.~2b), let us consider a locus of
the outermost cone containing 20 equi-spaced sparks at a distance $d=s_{out}\cdot
r_p=0.875\cdot 110 m\approx 100$~m from the magnetic pole. 
Both the HPBW diameter of sparks as well as distance between HPBW points is 
about $h=r_p/a\sim 17$~m. The line-of-sight
grazes this cone in such a way that the observer can detect up to three
spark-associated subpulses in a single pulse.


\vskip 12pt~\rotatebox{270}{\scalebox{.8}{\includegraphics[scale=0.8]{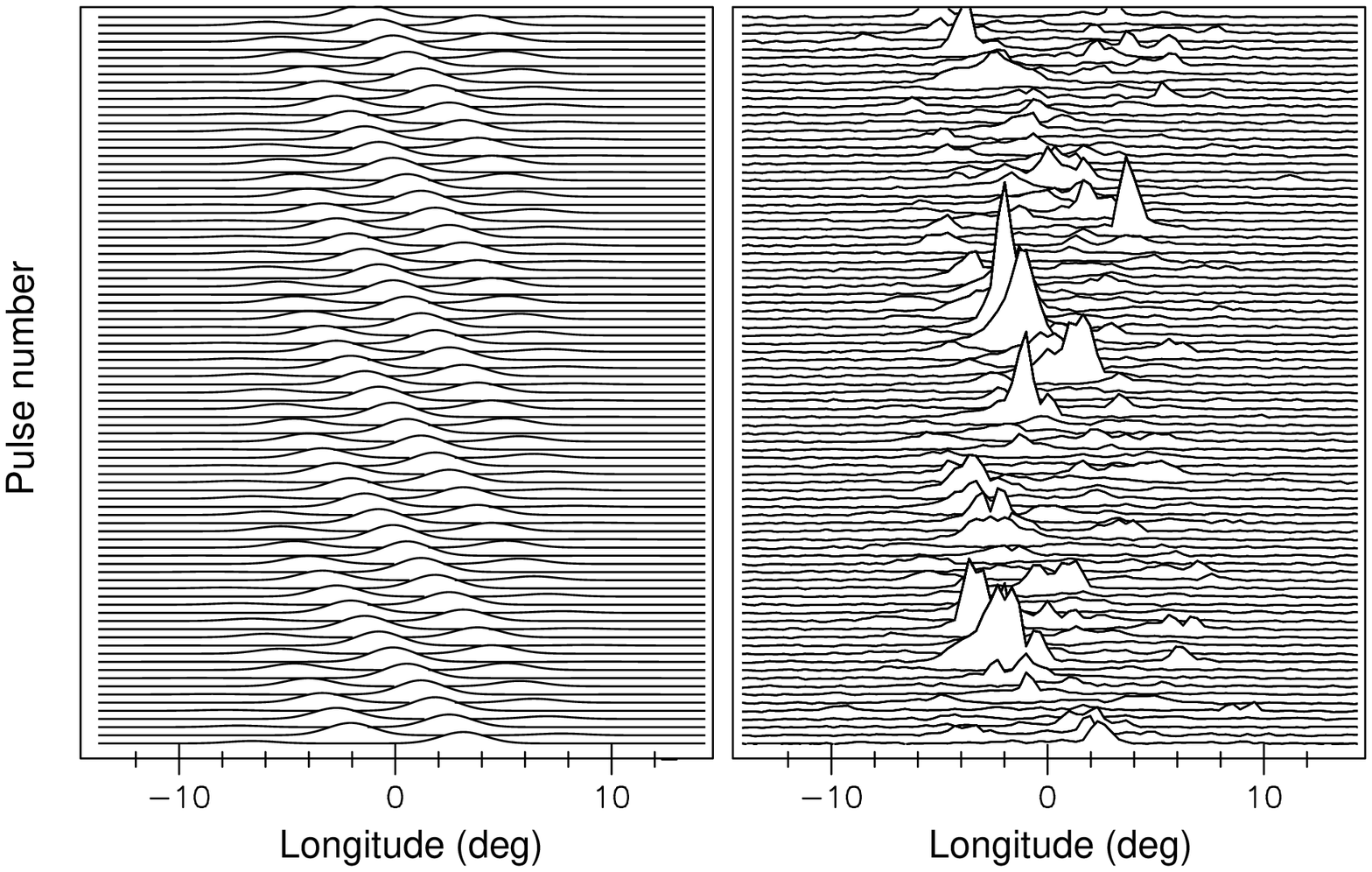}}}~\figcaption[fig2.eps]
{{\bf a}. Simulated subpulse drift patter for PSR B0943+10 
(left-hand side) and ---{\bf b}. observed pattern (right-hand side) after
Deshpande \& Rankin (1999). The apparent secondary drift-bands consist of subpulses belonging to every other pulse period.
\label{fig2}}~\vskip 2pt

For $P=1.1$~s and timing age $\tau_6=4.9$ milion years we have emission altitude at 430 MHz 
$r_6=r_{em}/R=55\cdot\nu_{GHz}^{-0.21}\tau_6^{-0.1}P^{0.33}\approx 60$ (Kijak and Gil 1997, 1998). 
Given the grazing sight-line geometry, we
can reasonably assume that the opening angle of the beam of this pulsar at
430 MHz is about the impact angle. Thus $|\beta|\stackrel{<}{\sim}%
\rho(0.43)=1^\circ.24~r_6^{1/2}(0.43)\cdot P^{-1/2}$, which for $r_6\approx
60$ gives $|\beta|\stackrel{<}{\sim}9^\circ$. On the other hand, 
from the polarization
measurements  we have $\sin\alpha/\sin\beta=3$ \citep{dr99}
and thus $\alpha\approx{\rm arcsin}(3\cdot\sin 9^\circ)=28^\circ$.

Assuming $\alpha=28^\circ$ and $|\beta|=9^\circ$ as initial values we
will now perform calculations of the radiation emitted tangent to the
dipolar field lines at altitude $r_{em}(0.43)=60\cdot R=6\cdot 10^7$~cm,
corresponding to a system of gaussian sparks presented in Fig.~1.
Each spark
circulates uniformly around the central polar spark (shadowed) at a rate of
about 9.7 degrees per pulsar period $(360^\circ/37)$. The resulting single
pulse emission pattern is presented in Fig.~2a, in comparison with the 430
MHz observed pattern presented by Deshpande \& Rankin (1999) in their
Fig.~1, and reproduced here in Fig.~2b. We have obtained a ``perfect match'' by
fine tuning geometrical parameters, which finally turned out to be 
$\alpha=26^\circ$ and $\beta=-8^\circ$. One can clearly see eight secondary
drift bands corresponding to apparent even-odd modulation, which is exactly
the sequence presented  by Deshpande \& Rankin (1999). The subpulses along
secondary drift bands belong to every other pulse. The primary drift bands
are not visible because subpulses drift very fast across the profile from
one pulse to another. Less than two pulsar periods $(P_3=1.87~P/{\rm cycle})$
are needed for the subpulse to reappear at the same longitudinal position,
producing an apparent even-odd modulation effect (secondary drift bands).
This effect is so specific that it gives full solution (drift pattern and
$\alpha, \beta$ values) in the simulation/fit procedure. Let us finally check 
for consistency that
${\cal N}=\hat{P}_3/P_3=37/1.85\sim 20$ (eqs. [\ref{pehat}] and [\ref{N1}]).

\subsection{PSR 2303+30}

This is another pulsar showing precise intensity modulation along secondary
dirft-bands, formed from even and odd pulses separately. Individual pulses
alternate between single-subpulse and double-subpulse form,
with astonishing precision. The primary subpulse drift rate is so fast that subpulses drift across the pulse window
from the edge of the pulse profile to its centre during about one pulsar
period $P$. The apparent periodicity $P_3/P$ is about 1.9 \citep{so75}. Our aim is to determine this fundamental
periodicity $\hat{P}_3$ (eq. [\ref{pehat}]) and the number of drifting sparks ${\cal N}$ (eq. [\ref{N1}]) forming the 
apparent secondary bands visible in Fig.~3b.

The high signal-to-noise ratio 430 MHz single pulse data have been recorded at the
Arecibo observatory in August 1986 (Gil et al. 1992). A typical sequence of 150 single pulses
of PSR B2303+30 is presented in Fig.~3b. One can distinguish two different
types of phase-versus-intensity modulation in this sequence.
Type A, visible near the top, demonstrates constant phase and
intensity in both single and double pulses. Type A modulation is preceded
and followed by Type B in which both phase and intensity of subpulses are
modulated along secondary driftbands. The modulation is very precise with
subpulse intensity strongly depending on its phase. The subpulses at the
trailing edge are weak. The intensity gradually increases until it reaches a
maximum near the center of the pulse window. Then it gradually decreases
again toward the leading edge of the profile.

The single-pulse polarization characteristics of PSR 2303+30 are also very
interesting. The position angle variation is presented in lower panel of
Fig.~3b, which shows the position angle at the peak of subpulses
\citep{ghn92}. Filled
symbols represent an intensity weighted average which can be interpreted as
the Rotating Vector Model (RVM) curve (Radhakrishnan \& Cooke 1969).

Thus, the intensity and polarization characteristics of PSR 2303+30
described above, are consistent with spark-associated subpulse beams
rotating around the magnetic axis (see Gil et al. 1995 for
discussion of polarization signatures of drifting subpulses). To simulate
its emission, we note that the complexity parameter $a$ (eq.~[\ref{n}]) in this
case $(P=1.57~s, \dot{P}_{-15}=2.9)$ is about 5. This implies that there are two nested cones
around the core beam
($a=5$, $n=2$; eqs.~[\ref{n}] and [\ref{a}]) Geometrical model of such a polar cap shows that $s_{out}\approx 0.85$, with 12
sparks forming the outer cone. Since the apparent period $P_3\approx 1.9$ \citep{so75}, thus according to
equation (\ref{N1}), the circulation period (eq.~[\ref{pehat}]) $\hat{P}%
_3\approx 23\cdot P$, that is each spark completes one full rotation around
the pole in about 23 periods $P=1.57$~s. This corresponds to the circulation
drift rate $D=360^\circ/23\approx 15.^\circ 6$ per period $P$.


\vskip 12pt~\scalebox{.8}{\includegraphics[scale=0.8]{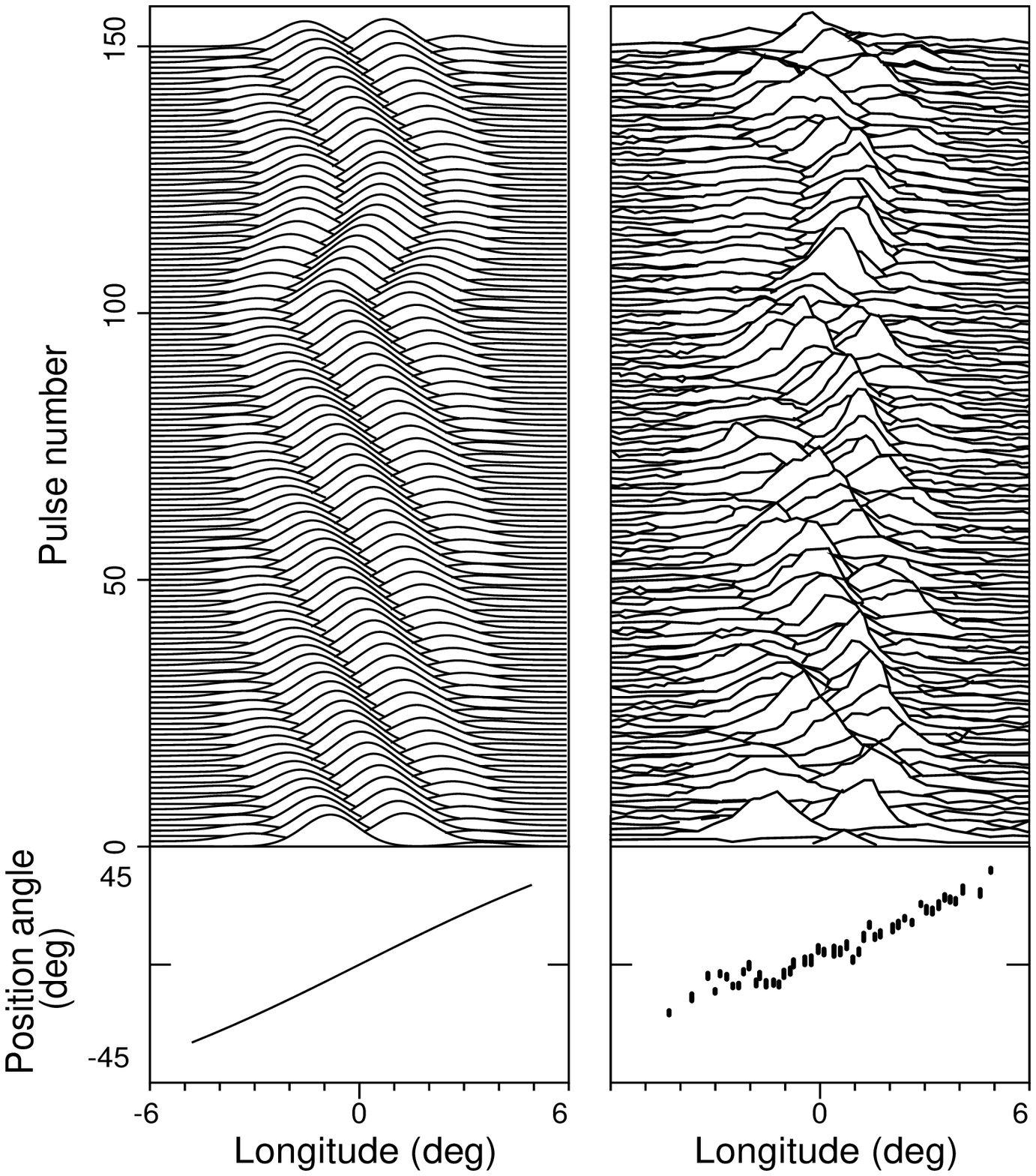}}~\figcaption[fig3.eps]
{{\bf a}. Simulated subpulse drift pattern for PSR B2303+30 (upper
left-hand side) and ---{\bf b}. observed pattern (upper right-hand side) observed at 
430 MHz from the Arecibo Observatory. Lower 
panels represent position angle variations: Rotating Vector Model for $\alpha=50^\circ$ and
$\beta=-6.^\circ 7$ (left-hand side), and position angle values measured at subpulse peaks (right-hand side).
Note that the apparent secondary drift-bands consist of subpulses belonging to every other pulse period.
\label{fig3}}~\vskip 2pt

We will now calculate the sequence of the first 100 single pulses of the
type $B$, using the value of $D=15.^\circ 6/P$ estimated above. The emission
altitude at 430 MHz is $r_6(0.43)\approx 60$ with $\tau_6=8.63$ (Kijak and Gil 1997, 1998). We
have found a number of secondary drift bands matching very well the observed pattern
for $\alpha=50^\circ$ and $\beta=-6.^\circ 7$. These values can be used to
estimate the position angle $\Psi$ swing within the RVM model (Radhakrishnan
\& Cooke 1969). In fact, $|d\Psi/d\varphi|=\sin\alpha/\sin\beta=6.5$, which
gives $\Delta\Psi\approx 52^\circ$ accross $\Delta\varphi\approx 8.5$ degrees
of longitude occupied by the pulse window. This is exactly what is observed
(compare lower panels of figs. 3a and 3b), confirming again self-consistency
of the model. 


\vskip 12pt~\scalebox{.65}{\includegraphics{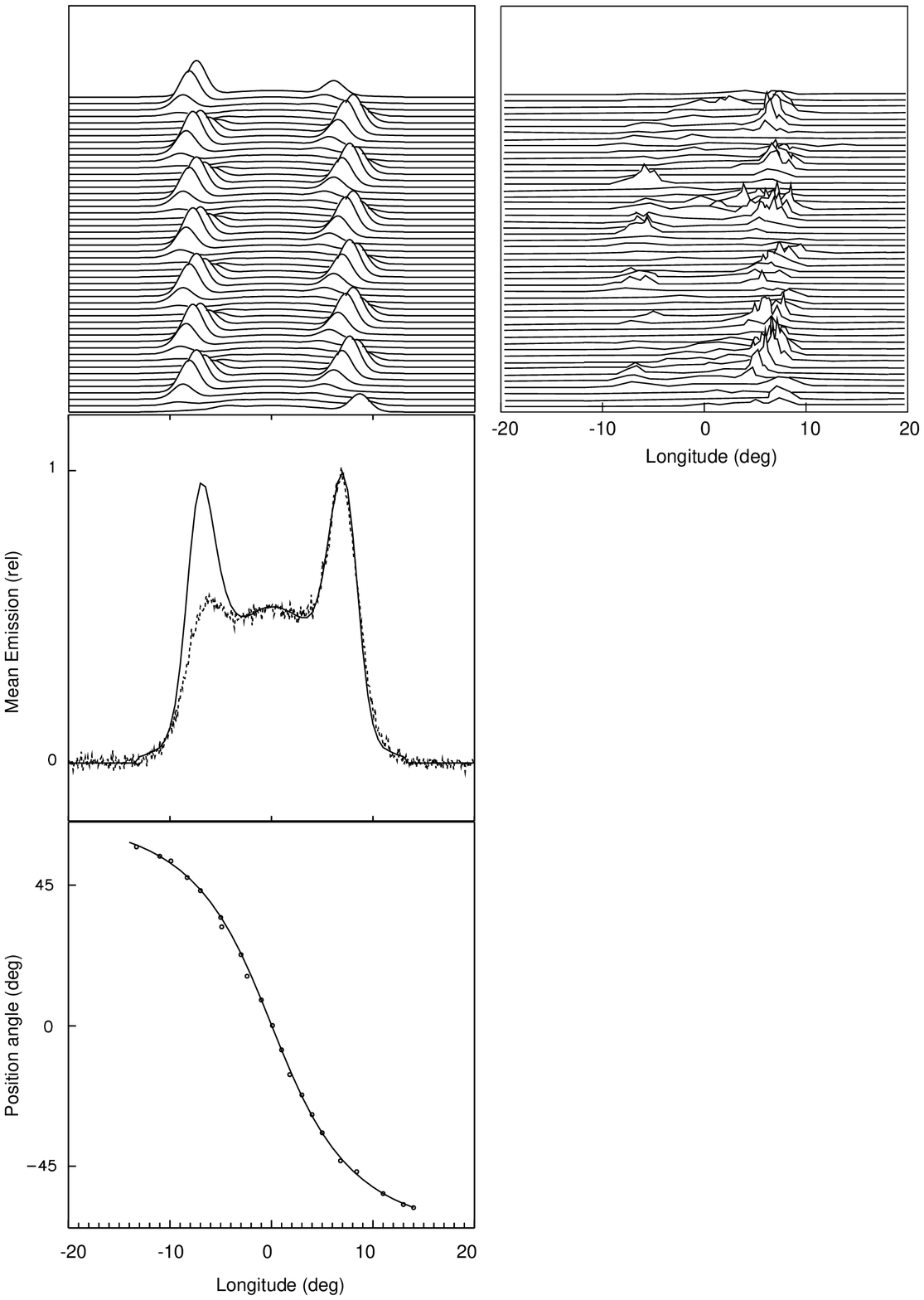}}~\figcaption[fig4.eps]
{{\bf a} Simulation of single pulses (left top panel), 
average profiles of both observational (dashed line) and simulated
(solid line) data, and position angle variation (lower panel) of the observed data
(circles) and model calculations (solid line) of PSR B2319+60.
---{\bf b} Sequence of 50 single pulses of PSR B2319+60 observed at 
1.4 GHz in the Effelsberg Observatory (right panel), to be compared with simulated pulses in the left panel. 
\label{fig4}}~\vskip 2pt

Above the pulse number 100, the secondary drift pattern changes into type
$A$, first reversing a drift direction and then keeping a constant phase for
about 20 periods. This corresponds to a slight change (less than 10\%) in
the circulation drift rate $D=14.^\circ 35/P$ and $14.^\circ 97/P$,
respectively, as compared with $15.^\circ 6/P$ in the type $B$ below and
above. Since $D\propto\Delta E$ (eqs. [\ref{vperp}] and [\ref{pehat}]), a
dramatic change in the apparent secondary drift pattern from type $B$ to
type $A$ corresponds to only few percent change in gap electric field.

\subsection{PSR 2319+60}

This is another pulsar demonstrating a very specific pattern of drifting
subpulses. The single pulse data at 1.4 GHz have been collected during the
pulsar month (September 1991) at the Effelsberg Observatory (Fig.~4)
and the average polarization data have been extracted from the Jodrell Bank
data base (Gould 1994). Since we have both a sequence of single
pulse data and the average profile, we can obtain the values of the
inclination $\alpha$ and the impact angle $\beta$ in the procedure of fitting actual
and simulated data. For $P=2.26$~s and $\dot{P}_{-15}=7$ we obtain $%
a=5.2$ and $n=2$ (eqs. [\ref{a}] and [\ref{n}]). 
As before, the observed drifting subpulses belong to the
second, outermost cone. The emission altitude at 1.4 GHz for $\tau_6=5$ is $%
r_6(1.4)=57$ and $\rho_{out}\approx 5^\circ$ for $s_{out}=0.85$. As one can
see from Fig.~4, we reproduced quite precisely the sequence of
drifting subpulses for ${\cal N}=9$ sparks. 
To match the observed primary drift rate $1.^\circ 8/P$ we have to choose ${\cal F}\approx
0.12$ (eq.~[\ref{calef}]) and $\hat{P}_3=70 P$ (eq.~[\ref{pehat}]). The simultaneous fit of single pulses
and the average profile gives $\alpha=27^\circ$
and $\beta=3^\circ$. This implies the circulation rate $D\approx 6.^\circ 2/P$. More importantly, the position angle curve calculated for these
values of $\alpha$ and $\beta$ (solid line in Fig.~4) matches perfectly the
observed values (dots), again confirming self-consistency of the
model. Let us note that $P_3/P=\hat{P}_3/({\cal N}P)=7.8$, close to observed value (Fig.~4).

\subsection{PSR B0031-07}

Vivekanand \& Joshi (1999) have recently published a sequence of drifting
subpulses of PSR B0031-07, which we reproduce here in Fig.~5b. The authors
report that the pair of drifting subpulses appears to be well separated and
that their amplitudes are anti-correlated with each other. On rare occasions,
three simultaneous subpulses seem to occur within one pulse. We
demonstrate that this behaviour is quite natural and show the sequence of
simulated single pulses (Fig.~5a) which match quite well the observed data
(Fig.~5b).

Since for $P=0.94$~s and $\dot{P}_{15}=0.4$ the complexity parameter $%
a\approx 4$, the number of cones is either one or two (eqs.~[\ref{n}] and [\ref{a}]). Let
us examine the model with $n=1$. The computer simulation gives ${\cal N}=5$ sparks
at $s\approx 0.7$. Thus for ${\cal F}\sim 0.1$, $\hat{P}_3\approx 33.65\cdot P$ and $P_3/P=34/5\sim 6.8$
(eq. [\ref{N}]), close to observed value (Fig.~5b). 
The model presented in Fig.~5a reproduces
the observed sequence of drifting subpulses quite well (if one ignores the
flux irregularities). In particular, one can see an anti-correlation of amplitudes
and occasional three subpulses within one pulse. The calculations are
related to frequency 327 MHz (Vivekanand \& Joshi 1999), which implies the
emission altitude $r_6=r_{em}/R\approx 48$ (Kijak \& Gil 1997, 1998). The
observing geometry for one ring with five sparks gives $\alpha=13^\circ$
and $\beta=4^\circ 0$. The model with two rings requires 11 sparks on the outer ring, leading to
$P_3=\hat{P}_3/N=34P/11\sim 3P$, which is inconsistent with observations. 


\vskip 12pt~\rotatebox{-90}{\scalebox{.8}{\includegraphics[scale=0.8]{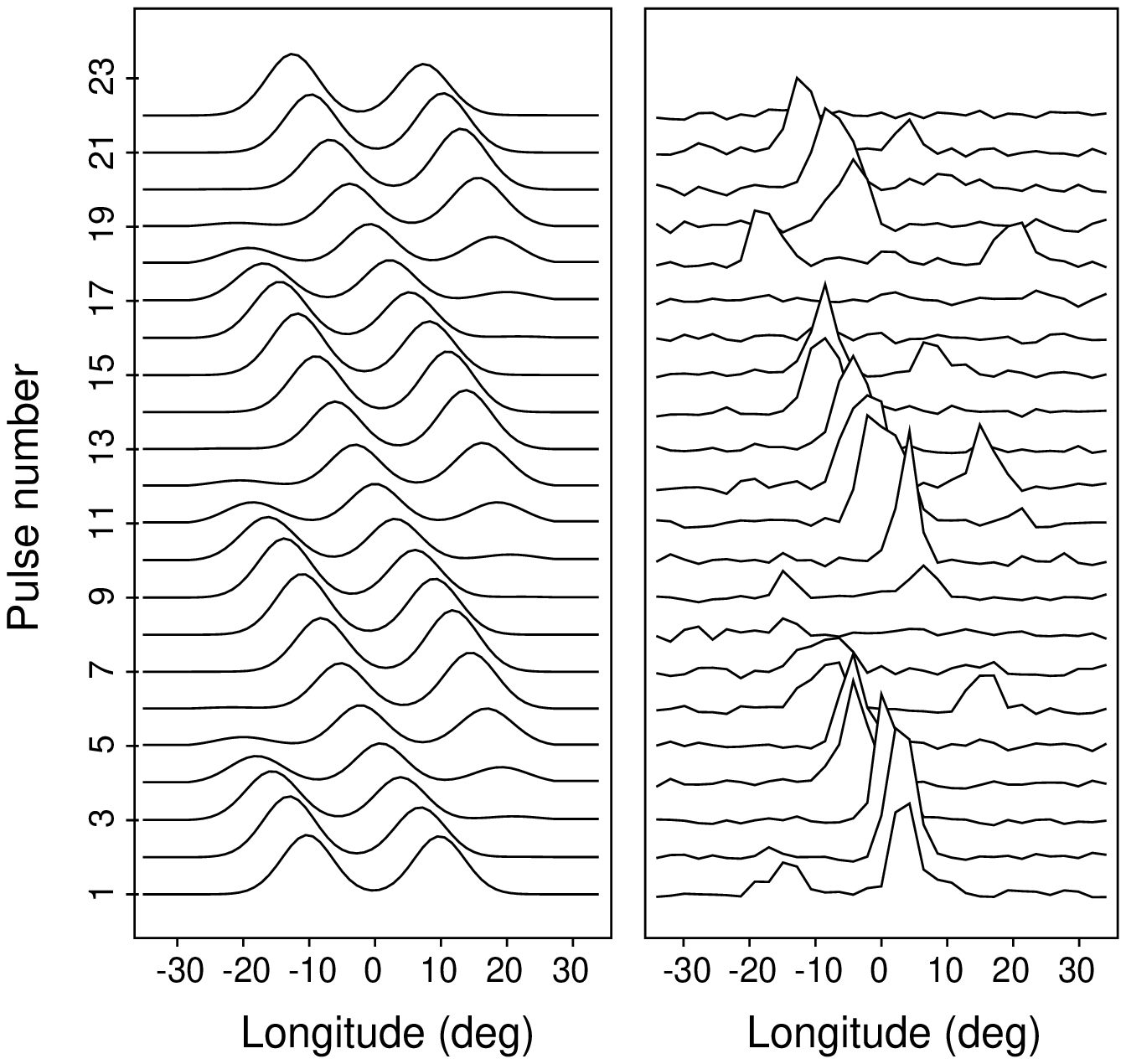}}}~
\figcaption[fig5.eps]{{\bf a}. Simulated subpulse drift pattern for PSR B0031$-$07 
(left-hand side) and ---{\bf b}. observed pattern (right-hand side) after
Vivekanand \& Joshi (1999). Note an occasional three subpulses within one pulse.
\label{fig5}}~\vskip 2pt

\section{Profile classification and ${\bf P}-\dot{\bf P}$ diagram}

Using equations~(\ref{erdepe}), (\ref{erpe}) and (\ref{a})
 we can calculate the period derivative (in
units of $10^{-15}$~s/s)
\be
\dot{P}_{-15}\approx 3\cdot 10^{-2}{\cal R}_6a^{3.5}P^{2.25} ,\label{pedot}
\ee
where the value of the complexity parameter $a\approx r_p/h$ 
can be obtained from the observed number of profile
components (eq.~[\ref{N}]) or more precisely from Rankin's profile classification
scheme (see below) and from the assumed range of ${\cal R}_6\sim 0.1$
(see Appendix).
Figure~6 presents the $P-\dot{P}$ diagram for 539 pulsars, out of which is 182
clasified within Rankin's scheme. 
For typical pulsars we follow classification established by Rankin
(1983, 1986, 1990, 1993a, b) except profile types determined after 1993 in
which case we follow Gould (1994). For millisecond pulsars we use Manchester
\& Johnston (1995), Xilouris \& Kramer (1996), Kramer et al. (1998) and
Xilouris et al. (1998).

For 182 classified pulsars, the value of the complexity parameter $a$ is calculated
from the equation~(\ref{a}) and represented in Fig.~6 by circle of different size (and colour corresponding to profile type). 
We excluded, for clarity of presentation, the three youngest
pulsars with $a>100$, so $1<a<100$ in Fig.~6. It is
interesting that no pulsar with $a\stackrel{<}{\sim}1$ exists. This means
that we can observe only those pulsars in which a mean free path for
$\gamma-B$ pair production is shorter than the size of the polar cap
($h<r_p$).  Another interesting fact is
that the lines of constant complexity parameter $a$ seem to follow a slope of 2.25, as can be
expected from equation~(\ref{pedot}). 

Rankin (1983, 1986) first proposed that the mean pulsar beam is arranged
into a core beam surrounded by two conal beams. Her proposal was
based on a careful analysis of morphological, polarization and spectral
properties of different profile components. It received further support by
calculations of the opening angles of different beams from the measurements
of pulse widths (Rankin 1990; 1993 a, b; Gil, Kijak \& Seiradakis 1993;
Kramer 1994 and Kramer et al. 1994). According
to Rankin's
classification scheme (see below) different profile species correspond to
different cuts through a nested multiconal pulsar beam. It is possible to
distinguish the core component from the conal ones not only by its location
within a pulse window (although generally the core component is flanked by
one, two or even three pairs of conal components), but also by its different
mean polarization and spectral characteristics as well as modulation
properties of corresponding single pulses. The conal components usually
show a relatively high degree of linear polarization with a regular swing of
the mean position angle following the so-called Rotating Vector Model (RVM,
Radhakrishnan \& Cooke 1969), but typically weak circular polarization. The
subpulses corresponding to conal components often show orderly subpulse
drift or at least periodic intensity modulations. On the other hand, the
subpulses in core components are rather longitude-stationary, i.e. they 
show no
apparent subpulse drift or strong intensity modulation. The circular
polarization in core components is quite high, often reversing sense at or
near the phase of maximum intensity. The linear polarization position angle
curve does not typically follow the RVM. Moreover, the spectra of core
components are steeper than those of conal components, meaning that the
former are more prominent at low radio frequencies while the latter dominate
at higher frequencies.

By analyzing the above properties in a large number of pulsars Rankin (1983)
distinguished seven major categories of pulse profiles: Multiple, Conal-triple, Quadruple, 
Triple, Conal-double, Conal-single and Core-single. We
have marked different profile species by different colours in Fig.~6. One can
immediately notice a tendency for grouping of different profile types in
different regions of the $P-\dot{P}$ diagram. The separation seems to follow
$P^{2.25}$ slope lines, corresponding to different values of the complexity parameter $a$, 
as suggested by equation (\ref{pedot}). We will discuss this
intriguing fact in the light of our spark-related, core/nested-conal model
of
the mean pulsar beam. We calculate the values of period derivative
$\dot{P}_{-15}$ from equation~(\ref{pedot}), using estimates of the parameter $a\approx
2n+1$ based on the number of cones required in different profile categories,
and then compare the results with the $P-\dot{P}$ diagram (Fig. 6) in
which the
complexity parameter $a$ (represented by the size of the circles) is
calculated from the equation~(\ref{a}), using only the basic pulsar parameters $P$ and
$\dot{P}$.


\vskip 12pt~\scalebox{.8}{\includegraphics[scale=0.85]{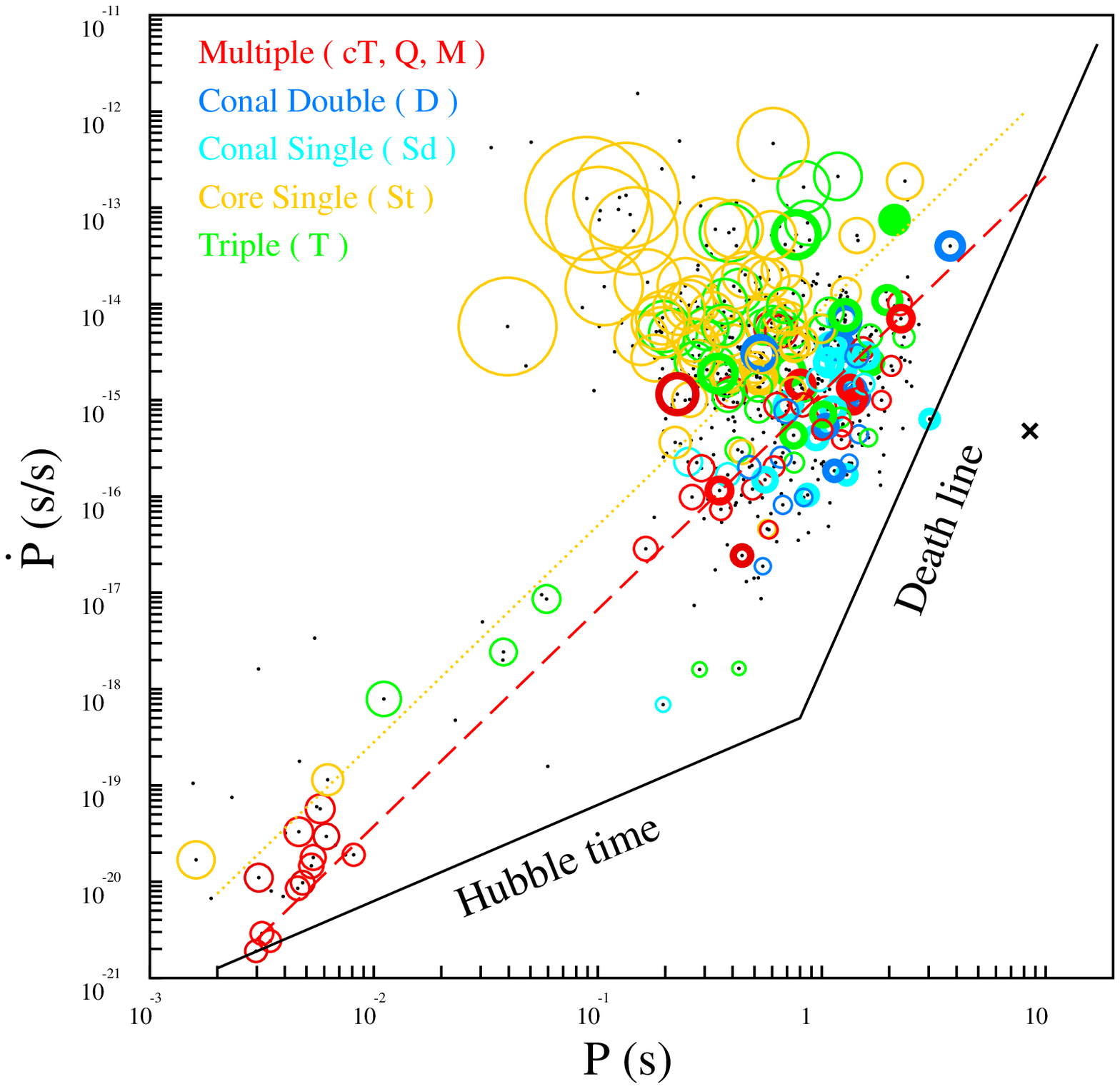}}~
\figcaption[fig6.eps]{$P-\dot{P}$ diagram for 539 pulsars with known (positive)
period derivative (dots). In 182 pulsars for which profile classification
is established (see text) we mark by different size circles the values of the complexity parameter $a=r_p/h$ calculated from
equation~(\ref{a}), and profile class is marked  by different
colours. The values of $a$ lie between about 1 (smallest circles) and 100
(largest circles) and we omit three youngest
pulsars with $a>100$ for scaling convienience. The number of sparks operating 
at any instant on the polar cap is approximately $a^2$, that is from one spark 
in the right lower corner to about ten thousands in the left upper corner of diagram. 
The thicker circles indicate
drifting subpulses and the green filled circle corresponds to the PSR
2002+31. The two lines with the $P^{2.25}$ slope
(eq.~[\ref{pedot}] with ${\cal R}_6=0.1$), correspond
to $a=5$ (multiple profiles - red) and $a=10$ (lower
limit for Core-single profiles - yellow), respectively.
PSR J2144-3933 is marked by the cross. The Hubble time and conventional death line are also marked.
\label{fig6}}~\vskip 2pt

\subsection{Multiple (red)}

We will include in this class the Conal-triple (cT - three conal
components), The Quadruple (Q - four conal components) and the Multiple (M
- core component flanked by two pairs of conal components) profiles, since
they all require $n=2$ conal beams\footnote{One can consider
three or even four nested cones in exceptional
cases (Gil \& Krawczyk 1997, Kramer 
1994, Gangadhara et al. 1999) 
but larger number of cones
cannot be clearly resolved into profile components ($a>10$, see
section
5.3 for details). Evidence for a large number of, barely resolved, profile
components is provided by Manchester et al. (1998) in their high
resolution
polarimetry of southern pulsars.} 
corresponding to $a\approx 2n+1=5$ (eq.~[\ref{a}]). 
The specific type of a profile depends on the observer's
line-of-sight: M - cutting both the core and two conal beams, Q - cutting
two cones but missing the core beam and cT - cutting the outer cone and
grazing the inner cone. 

Knowing the pulsar period and adopting the value of $a\approx 5$
for complex profile pulsars we can calculate the period
derivative (eq.~[\ref{pedot}]) to within an accuracy determined by the range of
${\cal
R}_6$ and compare it with the measured value. For multiple profile pulsars 
the following relationship should hold $P_{-15}\sim{\cal R}_6\cdot P^{2.25}$. 
It seems that this is really so. In fact, the red dashed line in Fig.~6 
corresponds to ${\cal R}_6=0.1$ (see Appendix). Already \citet{r92} has noticed that $M$-type pulsars
with typical periods tend to follow a line with similar slope. Here we confirm this tendency for larger group
of pulsars, including the millisecond ones.

\subsection{Conal Double (dark blue) and Conal Single (light blue)}

The Conal double profiles ($D$) have two conal components with no evidence
of core emission in the saddle between them. While the $D$ profiles are
double in the entire pulsar spectrum, the Core single ($S_d$) profiles
bifurcate only at low radio frequencies, clearly indicating that the
observer is grazing the conal beam with the opening angle increasing as 
frequency decreases. In Conal Double profiles the observer's impact angle is
smaller but in both cases the core beam is missing. In principle, one should
consider one or two cones around the core beam, meaning that $n=1$ or $n=2$
and thus $a=3$ or $a=5$. Therefore, both $D$ and $S_d$ profile pulsars
should occupy the common area with Multiple type pulsars. Figure 6 clearly shows
that this is really so.

\subsection{Core Single (yellow)}

The Core single ($S_t$) profiles are simple (almost Gaussian) in shape, with
prominent, sense-reversing circular polarization and rapid, non-RVM, position
angle swings. No drifting subpulses have been found in $S_t$ pulsars. More
generally, the emission of $S_t$ pulsars seems amorphous, with no hints of
subpulse modulation in their single pulses. In principle, one has to
consider two possibilities:

\begin{description}
\item (a) Just one (polar) spark with dimension comparable to the polar cap
size. In such a case $r_p\sim h$, meaning that a mean-free path for
$\gamma-B$ pair production is barely satisfied. It is no wonder then that $S_t$
pulsars do not seem to occur in the region of small $a\approx 1$ in Fig.~6.
\item (b) A very large number of small sparks occupy the polar cap. This means that radiation
of the adjacent sparks overlap and cannot be resolved into subpulses
(components). Since the radiation is tangent to dipolar magnetic field lines
to within an angle $1/\gamma$, where $\gamma$ is the Lorentz factor of the
emitting source, we have the condition $\Delta\rho<1/\gamma$. This implies that
the difference of the opening angles\footnote{The opening angle $\rho$ is
an angle between the magnetic dipole axis (pulsar beam axis) and tangent
to dipole field line corresponding to a particular emission feature.}$\rho$ of 
field lines corresponding to adjacent sparks is smaller than the angular
extent of elementary relativistic
emission. For dipolar field lines, the opening angle $\rho\approx
10^{-2}\cdot(d/r_p)\cdot(r/R)^{1/2}P^{-1/2}$, where $d<r_p$ is the distance
from
the pole to the base of the field line on the polar cap and $r=r_6/R$ is
the
emission altitude. Thus, $\Delta\rho\approx
10^{-2}({\cal D}/r_p)r_6^{1/2}P^{-1/2}$, where ${\cal D}=\Delta d\approx h$ is a characteristic
spark dimension and a typical distance between sparks. Kijak \& Gil (1997,
1998) demonstrated that the emission altitude is apparently period dependent
in such a way that $r_6^{1/2}P^{-1/2}\approx 10$. Since ${\cal D}/r_p\approx
h/r_p=1/a$ (eqs.~[9] and [12]), we have $\Delta\rho=0.1/a<1/\gamma$ and if
$\gamma$ is about 100 we conclude that $a>10$ for the $S_t$
profile
pulsars. Thus, the Core Single pulsars should  lie predominantly above the yellow
dotted line in Fig.~6, corresponding to $a=10$ calculated from equation~(\ref{a}).
\end{description}

The above consideration strongly suggests that the maximum
number of nested cones (eq.~[\ref{n}]) that can be resolved as profile components is about 4.
Interestingly, Mitra \& Deshpande (1999) have recently found, on completely
different grounds, that the number of nested cones within the overall pulsar
beam is not larger than 4. 
Therefore, according to equations (\ref{n}), (\ref{a}) and (\ref{N}), the  number
of resolved profile componets should not excced 9, which seems to be confirmed
observationally. Few millisecond pulsars were reported to show more than
7 profile componets (Kramer et al. 1998), which was a result of including
complex interpulses, as well as pre- and post-cursors, in the component count.
However, one cannot exclude even nine-component profiles for which $a=9$, and such
a case was recently reported by Gangadhara et al. 1999.
 
It is well known that about 65\% of $S_t$ profile pulsars develop weak
conal outriders at high frequencies, where the core emission is no longer a
dominant component of pulsar emission. We propose an explanation of this
phenomenon which at
first sight may seem a bit speculative, but a closer look shows that it can be
quite plausible. The gap height $h$ (which determines in our model 
both a spark dimension and a distance between sparks) depends on the
magnetic field like $B^{-4/7}$. In a multipolar surface magnetic
field it can happen that the value of $B$ drops by a factor of several to
ten from the local pole to the polar cap edge. In such a case the side
sparks would be a few times larger than the polar ones. If the
condition
$a=r_p/h>0.1\gamma$ is not satisfied at the polar cap boundaries, the side
sparks can form an outer cone of emission surrounding the core beam
(resulting from unresolved emission of a large number of much smaller sparks
well inside the polar cap). One cannot exclude that an undetectable outer
beam always exists in the $S_t$ profile pulsars and therefore the core
emission of these objects represents only the inner part of the polar cap.


\vskip 12pt~\scalebox{.75}{\includegraphics[scale=0.75]{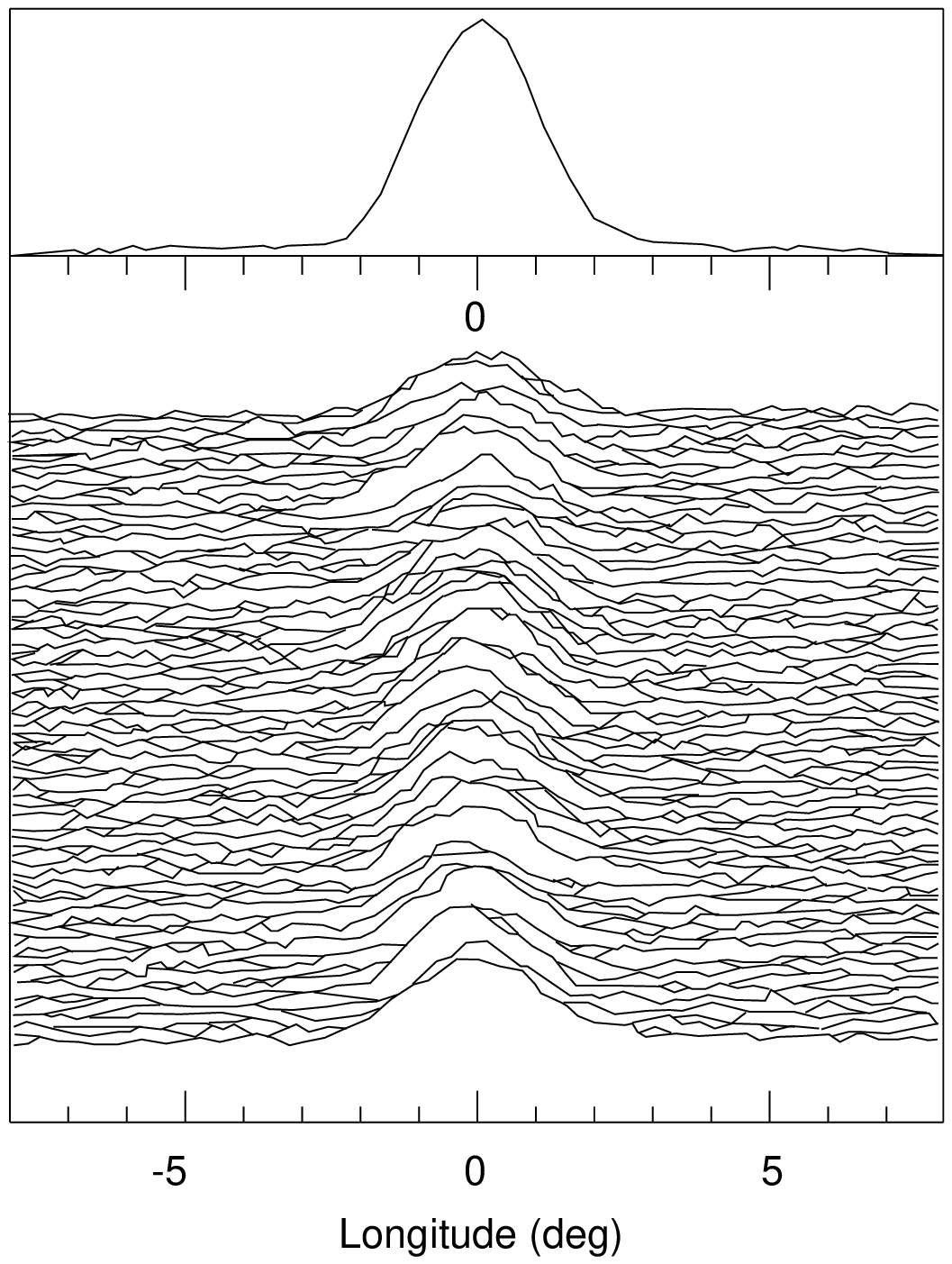}}~
\figcaption[fig7.eps]{430 MHz total intensity observational data of PSR 2002+31
taken at 430 MHz in the Arecibo Observatory. Note that both the mean
profile (top) and individual pulses of this pulsar have similar Gaussian
shapes. The position of PSR B2002+31 in Fig.~6 is marked by filled
symbol.
\label{fig7}}~\vskip 2pt

\subsection{Triple (green)}

A genuine Triple (T) profile consists of a central core component flanked by a
pair of conal outriders in the entire pulsar spectrum (although the core
component may sometimes dissappear at high frequencies). Here again we have
two possibilities:

\begin{description}
\item (a) Just one cone around the core beam and an almost central
line-of-sight
trajectory. This case corresponds to $n=1$ and $a=3$ (eqs.~[\ref{n}] and[\ref{a}]) and thus 
$T$ profile pulsars should occupy a region on the $P-\dot{P}$ diagram 
corresponding to $a^{3.5}\sim 50$ in equation (\ref{pedot}).
\item (b) A large number of small sparks corresponding to $a>10$, with the
outer ring generated in the same way as in the case of the $S_t$ profiles
with outriders (section 5.3). The only difference is that the conal
emission in $T$
profile pulsars is not dominated by the core emission even at low
frequencies, so they are of triple type in the entire pulsar radio spectrum.
Since $a^{3.5}>1000$ in equation (\ref{pedot}), these pulsars should lie
considerably above Multiple, Conal Double and Conal Single profile pulsars,
overlapping partially with Core Single profile pulsars. This is really observed
on the $P-\dot{P}$ diagram presented in Fig. 6.
\end{description}

A good example of T-profile pulsar with $a\geq 10$ is PSR
B2002+31. This pulsar with $P=2.11$~s and $\dot{P}_{-15}=75$ has the
complexity parameter $a=11$ and lies just above the yellow dotted line
corresponding to $a=10$ in Fig.~6 (filled green circle). The Gaussian
waveform (Fig.~7) in the profile centre (top) consist of almost Gaussian individual
pulses (each, in our interpretation, corresponding to a large number of
``$1/\gamma$ unresolved'' sparks). The flat profile wings evolve into a pair of
conal outriders at higher frequencies above 1 GHz (e.g. Gould \& Lyne
1998, Table 4). 

\section{Discussion and Summary}

There is a growing evidence from pulsars with orderly drifting
subpulses \citep{dr99,vj99} that the RS75-type polar gap does exist \citep{xqz99}. 
Analyzing observational properties of both single pulses and mean profiles we have found that the
height scale of inner gap should be close to the value given by RS75.
Although the physics behind our model is constrained only by dimensional analysis,
it is a useful empirical base for more detailed modeling and interpretation of observations.
We demonstrate that all periodicities associated with drifting subpulses are determined mainly by the $P$
and $\dot{P}$ values. If the observing geometry is known, then one can even reproduce patterns of drifting subpulses.
We found amazing agreement between simulated and observed drift-bands in a number of cases (Table 1).

Throughout this paper we assume and/or argue that:
\begin{description}
\item(1) At any given time the polar cap is populated as densely as
possible with a number of isolated spark discharges. This number is 
approximately equal to $a^2\sim 25\cdot\dot{P}_{-15}^{0.6}\cdot P^{-1.3}$.
\item(2) A spark characteristic dimension as well as the typical distance
between sparks is about the polar gap height $h\approx 3\cdot 10^3P^{1/7}
\dot{P}_{-15}^{-2/7}$~cm.
\item(3) The life-time (exponentiation time $\tau\approx {\cal R}/c$)
of each spark is of the order of 10 microseconds but they tend to reappear
in
almost the same places on time-scales shorter than $h/c\ll 10\mu$sec.
\item(4) The actual surface magnetic field is dominated by multipole
components. The planes of field lines tend to converge towards a local
pole, which does not coincide in general with the global dipole axis. 
We believe that the actual surface magnetic field has a ``sunspot-like'' structure with one pole
at the dipolar polar cap.
\item(5) One spark is anchored quasi-permanently (oscilating on $10\mu$s
time scale) to the local pole, while other sparks
perform a circumferential ${\bf E}\times {\bf B}$ drift around the polar spark.
\item(6) Sparks supply a corresponding subpulse-associated plasma columns
that can give rise to coherent radio emission due to some instability
developing in a purely dipolar magnetospheric region at altitudes of about few percent of the light-cylinder radius
$R_L=cP/2\pi$.
\item(7) The polar spark is associated with the core pulsar emission while
other sparks circulate around the pole and contribute to the conal pulsar emission. 
\end{description}

The assumptions and/or arguments used in this paper lead to nested-cones model of
the
mean pulsar beam. Such a beam structure has already been deduced by Rankin
(1983) as a result of her profile classification scheme. This has been
confirmed later by the analysis of the period dependence of pulse widths
of different profile species (Rankin 1993 a,b; Gil, Kijak \& Seiradakis
1993; Kramer et al. 1994; Gil \& Krawczyk 1996, Gil \& Han 1996, Mitra \&
Deshpande 1999). 
It is worth mentioning that our arguments are completely different from
those of Rankin. Therefore, the fact that our model predictions are in
excellent agreement with Rankin's classification scheme is not trivial.

The number of nested cones surrounding the core beam depends in our model
mainly on basic pulsar parameters $P$ and $\dot P$ (a dependence on the
unknown parameters characterizing the structure of the actual surface
magnetic field is rather weak). Therefore we can make three strong model
predictions:
\begin{description}
\item(a) The Core Single ($S_t$) and Conal Single ($S_d$) profile pulsars
should be well
separated on the $P,\dot{P}$ diagram (Fig.~6). A division line corresponds
to $a\approx r_p/h\sim 10$ (yellow dotted).
\item(b) The single pulse emission in $S_t$ pulsars
(with $a>10$) should be amorphous with no hints of modulation on the
subpulse time
scales, while $S_d$ pulsars should demonstrate subpulses in their single
pulses. 
\item(c) The subpulses in core components of complex profiles should be
longitude-stationary while an apparent subpulse drift or periodic intensity modulation should occur
exclusively in conal
components. 
\end{description}

These prediction, which seem to be supported observationaly,
are a direct consequence of a
non-stationary polar gap discharge through a number of 
isolated sparks,
reappearing in almost the same place for a long time as compared with
their lifetime. This is an important advantage of sparking models over
stationary free-flow models, where all the above observational
features have no natural, self-consistent explanation. Perhaps an even more
important advantage of a non-stationary polar gap discharge is the
generation of a two-stream plasma instability in the magnetosphere near
the neutron star (Usov 1987;  Asseo \&
Melikidze 1998). The sparking phenomenon creates a succession
of plasma clouds moving along magnetic field lines, each containing
particles with a large spread of momenta. Overlapping of particles with
different energies from successive clouds ignites strong Langmuir
oscilations, which may lead eventualy to the generation of coherent pulsar
radio emission (Melikidze, Gil \& Pataraya 1999). Interestingly, this
instability is the only one which, according to our knowledge, develops
at altitudes of the order of 1 percent of the light cylinder radius,
where
the
pulsar radio emission is expected to originate (Cordes 1978, 1992; Kijak \&
Gil 1997, 1998). However, the latter statement is again true only if the potential drop height
scale in the actual polar gap is close to that described by the RS75. 

In the accompanying paper Melikidze, Gil \& Pataraya (1999) examine a
non-linear evolution of Langmuir electrostatic oscillations in plasma clouds
associated with sparks reappearing at approximately the same place of the
polar cap, according to a model presented in this paper. They found that a well known 
modulational instability leads
to formation of a ``bunch-like'' charged solitons, capable of generating coherent
curvature radiation at radio wavelenghts. 
Thus, we believe that the non-stationary sparking discharge of the 
polar gap driven by nondipolar surface magnetic field 
explains not only observational characteristics of pulsar radiation modulation but
also the mysterious generation mechanism of this radiation, all in a
self-consistent way.

\vspace{.5cm} 
\ni {\bf Acknowledgements}
This paper is supported in part by 
the KBN Grant 2~P03D~015~12 of the Polish State Committee for Scientific
Research. 
We thank A.A. Deshpande, J.M. Rankin, M. Vivekanand and B.C. Joshi for
providing us raw data with drifting subpulses.
We also thank K.S. Cheng, T.H.~Hankins, A. Krawczyk, G. Melikidze, D. Mitra and
L. Nowakowski  for their valuable
contributions to this work. We thank F.C. Michel, D. Nice, 
B. Zhang and L. Zhang for helpful discussions
and especially Z. Arzoumanian, M. Kramer and D. Lorimer for critical reading of parts of the manuscript and many
useful comments. 
We also thank E. Gil for technical
assistance. The Arecibo Observatory is operated by Cornell University under a
Cooperative Agreement with the U.S. National Science Foundation.

\appendix

\section{Sunspot-like type surface magnetic field}

Radio pulsars are believed to turn off when they can no longer produce
electron-positron pairs in strong magnetic fields just above the polar cap.
The limiting rotational period $P$ at which this occurs depends on the 
magnitude and configuration of the surface magnetic field $B_{s}=b\cdot B_d$ (see eq. [\ref{erpe}]). 
Unfortunately, only the perpendicular component of dipolar field $B_{d}$ can be 
deduced from the observed spin-down rate 
$\dot{P}$, i.e. $B_d=3.2\cdot 10^{19}(P\cdot\dot{P})^{1/2}$~Gauss. The line on the $B_{d}-P$ 
plane or $\dot{P}-P$ plane corresponding to the critical period is called 
a death line. 
No radio pulsar should be observed to the right of this line i.e. 
with period longer
than the critical one. Recently, \citet{ymj99} reported the existence 
of PSR J2144$-$3933 with a period of 8.5 s, which is located to the right of all known death
lines. As \citet{ymj99}
conclude themselves, under the usual assumptions, this slowly rotating 
pulsar should not be emitting a radio beam.

Here we consider a death-line problem for PSR J2144$-$3933, assuming
our preferred configuration of surface magnetic
field, that is a sunspot type. Following \citet{cr93} we can write the death-line equation in the form
\be
\log B_d=1.9\log P-\log B_s+0.6\log{\cal R}+21 , 
\label{A1}
\ee
where we introduced radius of
curvature ${\cal R}$ of surface field lines as the unknown variable in their equation (9).
Setting parameters of PSR J2144-3933 $P=8.5$~s and $B_d=2\cdot 10^{12}$~G, we find that 
$B_s\stackrel{>}{\sim}10^{13}$~G and ${\cal R}<<10^6$~cm. 
In fact, for ${\cal R}=10^6$ cm the 
surface field $B_{s}=10^{14.06}$ G, which is greater than
the critical magnetic field $B_q=4.4 \cdot 10^{13}$ G above which 
the photon splitting phenomenon would quench the radio 
pulsar (Baring \& Harding, 1998). 
The entire manifold of marginal death lines following from the above equation
applied to PSR J2144-3933 is represented by the solid line in 
Fig.~8. All points $(B_s, {\cal R})$ lying on this line correspond
to death lines (eq. [\ref{A1}]) which pass through PSR J2144-3933 marked
by cross in Fig. 6. As one can see, the inferred radius of curvature ${\cal R}$ for 
strong field $B_s\gg B_d$ is about $10^5$~cm. We believe, that the existence of PSR J2144$-$3933 
is at least consistent, if not implicative, with a sunspot-like surface magnetic field
at pulsar polar cap.


\vskip 12pt~\rotatebox{-90}{\scalebox{.7}{\includegraphics[scale=0.7]{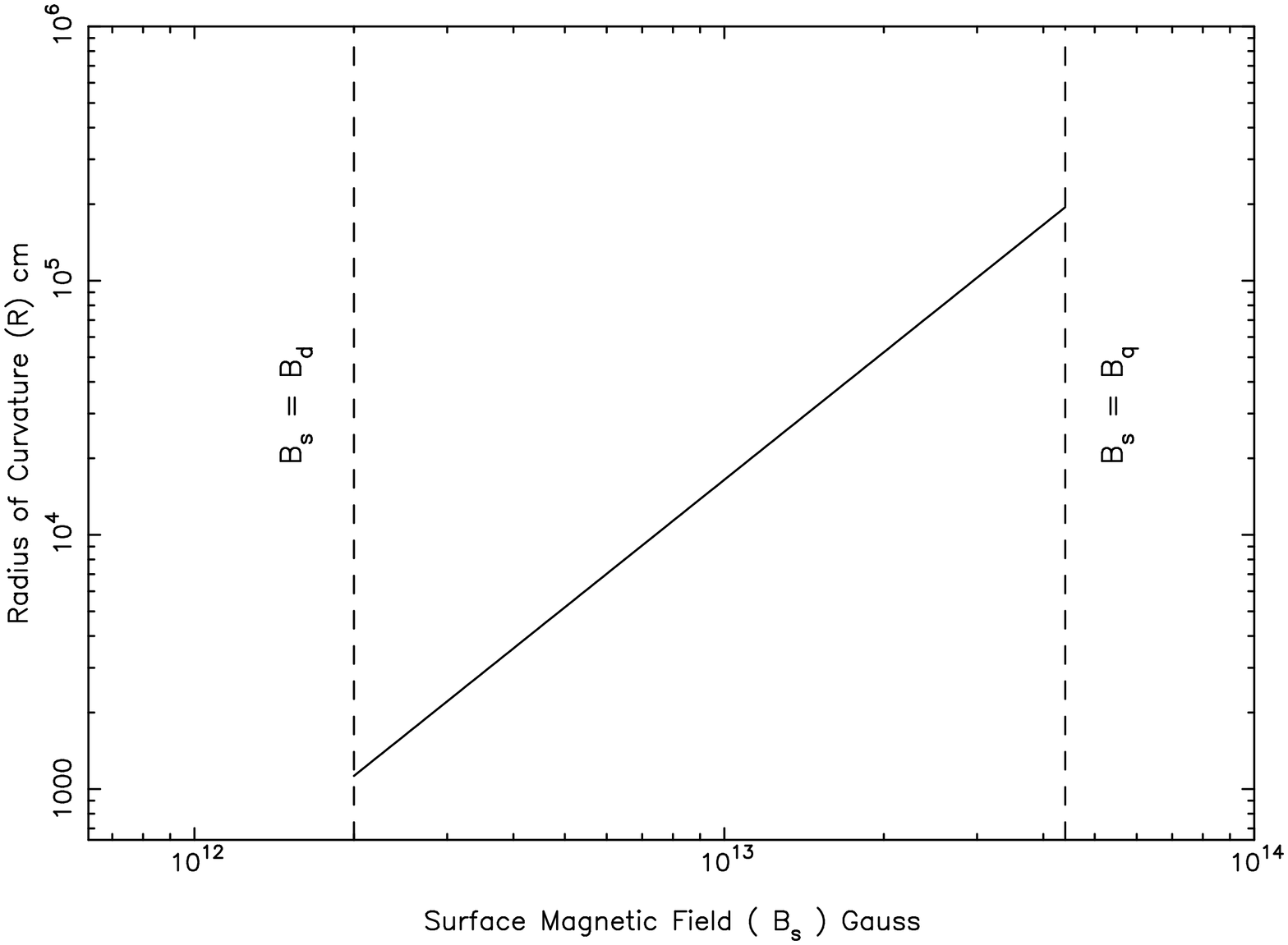}}}~
\figcaption[fig8.eps]{Radius of curvature ${\cal R}$ versus the magnitude $B_s$ of the putative sunspot surface magnetic
field in PSR J2144$-$3933 (eq. [\ref{A1}] for $P=8.5$~s and $B_d=2\cdot 10^{12}$~G). The two dashed vertical
lines correspond to $B_s=B_d=2\cdot 10^{12}$~G and $B_s=B_q=4.4\cdot 10^{13}$~G, respectively. The position of
PSR J2144$-$3933 in Fig.~6 is marked by the cross.
\label{fig8}}~\vskip 2pt


\clearpage
\begin{deluxetable}{llllllllllll}
\footnotesize
\tablewidth{13.5cm}
\tablecaption{}
\tablehead{
\colhead{PSR B} &\colhead{$P$[s]} & \colhead{$\dot{P}_{-15}$} & \colhead{$a$} & \colhead{$n$} & \colhead{${\cal N}$} & \colhead{$\frac{\hat{P}_3}{P}$} & \colhead{$\frac{P_3}{P}$} & \colhead{$s_{out}$} & \colhead{${\cal F}$} & \colhead{$\alpha $} & 
\colhead{$|\beta|$}} 
\startdata
0943$+$10 & 1.09 & 3.52 & 6.8 & 3 & 20 & 37 & 1.85 & 0.875 & 0.125 & 26 & 8 \\ 
2303$+$30 & 1.57 & 2.9 & 5 & 2 & 12 & 23 & 1.92 & 0.85 & 0.11 & 50 & 6.7 \\ 
2319$+$60 & 2.26 & 7 & 5.2 & 2 & 9 & 70 & 7.8 & 0.85 & 0.12 & 27 & 3 \\ 
0031$-$07 & 0.94 & 0.4 & 4 & 1 & 5 & 34 & 6.8 & 0.7 & 0.1 & 15 & 3.3 \\ 
\enddata
\tablecomments{Pulsar name, period $P$, period derivative $\dot{P}$, in units of
$10^{-15}$~s/s, complexity parameter $a$ (eq.~[\ref{a}]), number of nested cones
$n$ (eq.~[\ref{n}]), number of circulating beams ${\cal N}$ within outermost cone,
circulation period $\hat{P}_3$ (eq.~[\ref{pehat}]) in units of $P$, locus of the
outermost cone $s_{out}$ in units of the polar cap radius $r_p$, filling
factor ${\cal F}$ (eq.~[\ref{calef}]), inclination $\alpha$ and impact
$\beta$ (absolute value) angles in degrees.}
\end{deluxetable}
\end{document}